\documentclass[preprint,12pt]{elsarticle}
\usepackage[a4paper, left=2.5cm, right=2.5cm, top=3cm, bottom=3cm]{geometry}

\usepackage{amssymb}

\usepackage{preamble}
\usepackage{cleveref}

\begin{document}

\begin{frontmatter}




\title{Learning Physics-Consistent Material Behavior from Dynamic Displacements}


\author[inst1]{Zhichao Han}

\affiliation[inst1]{organization={Institute for Building Materials, ETH Z\"urich},
            addressline={Laura-Hezner-Weg 7}, 
            postcode={8093}, 
            state={Z\"urich},
            country={Switzerland}}

\author[inst1]{Mohit Pundir}
\author[inst2]{Olga Fink\fnref{fn1}}
\author[inst1]{David S. Kammer\corref{cor1} \fnref{fn1}}
\cortext[cor1]{Corresponding author. E-mail: \textsf{dkammer@ethz.ch}.}
\fntext[fn1]{Joint supervision.}

\affiliation[inst2]{organization={Laboratory of Intelligent Maintenance and Operations Systems},
            addressline={Station 18}, 
            postcode={1015}, 
            state={Lausanne},
            country={Switzerland}}

\begin{abstract}
Accurately modeling the mechanical behavior of materials is crucial for numerous engineering applications. The quality of these models depends directly on the accuracy of the constitutive law that defines the stress-strain relation. However, discovering these constitutive material laws remains a significant challenge, in particular when only material deformation data is available. To address this challenge, unsupervised machine learning methods have been proposed to learn the constitutive law from deformation data. Nonetheless, existing approaches have several limitations: they either fail to ensure that the learned constitutive relations are consistent with physical principles, or they rely on boundary force data for training which are unavailable in many in-situ scenarios. Here, we introduce a machine learning approach to learn physics-consistent constitutive relations solely from material deformation without boundary force information. This is achieved by considering a dynamic formulation rather than static equilibrium data and applying an input convex neural network (ICNN). We validate the effectiveness of the proposed method on a diverse range of hyperelastic material laws. We demonstrate that it is robust to a significant level of noise and that it converges to the ground truth with increasing data resolution. We also show that the model can be effectively trained using a displacement field from a subdomain of the test specimen and that the learned constitutive relation from one material sample is transferable to other samples with different geometries. The developed methodology provides an effective tool for discovering constitutive relations. It is, due to its design based on dynamics, particularly suited for applications to strain-rate-dependent materials and situations where constitutive laws need to be inferred from \emph{in-situ} measurements without access to global force data.
\end{abstract}




\end{frontmatter}



\section{Introduction}
\label{sec:introduction}
Understanding the constitutive relation of materials is of fundamental importance for modeling and predicting their mechanical behavior. Accurately modeling the material behavior benefits numerous applications, including reliable prediction of landslides in nature~\cite{cicoira2022towards}, inverse design of mechanical materials in science and engineering~\cite{bastek2023inverse}, generating realistic scenarios in movies~\cite{stomakhin2013material}, and many more. Since the constitutive relation cannot be measured directly in experiments, it is inferred indirectly through measurable and complementary quantities, such as macroscopic deformation and the corresponding mechanical response. The fundamental concept is to infer a surrogate constitutive relation with which the simulated behavior matches the true behavior~\cite{mottershead1993model}.  To this end, conventional approaches typically assume a user-defined functional form (based on well-known material models) for the constitutive relation. Experimental data, such as displacement or strain data combined with the applied/reaction forces, are then used to infer the parameters within these models~\cite{mihai2017family}. Various techniques such as strain gauges~\cite{harding1983tensile} and digital image correlation (DIC)~\cite{rubino2019full} are commonly used to collect such complementary data, enabling the use of sophisticated strategies such as finite element model updating (FEMU)~\cite{marwala2010finite} to calibrate parameters in the pre-assumed material model. Despite its usefulness, selecting an appropriate functional form for the constitutive relationship is challenging. It requires a high level of expertise and can often result in an oversimplified representation of the actual material behavior~\cite {balmforth2014yielding, kolymbas2012constitutive,mihai2017family}. 
The challenge to infer the constitutive model is further exacerbated in \emph{in-situ} measurements where only one of the complementary data (either strain or boundary force but not both) is accessible. For example, measuring the biomechanical response (force measurement) \emph{in-situ} of biological tissue to identify lesions and diseases is particularly difficult~\cite{lim2009situ, mazza2008constitutive}. 
In addition, understanding the material behavior of rate-dependent materials is crucial for applications such as simulating seismic site response~\cite{miao2022reproducing} or analyzing the behavior of certain polymer materials~\cite{siviour2016high, field2004review}. This requires consideration of dynamic deformation, where the constitutive relation depends on high strain rates. Consequently, determining constitutive relations without predefined functional forms and under various loading conditions (static or dynamic) remains a challenging but impactful task.


Machine learning (ML), particularly neural-network-based approaches, offers a promising alternative for acquiring constitutive relations~\cite{fuhg2024review}. These methods replace user-defined function forms with neural networks, reducing the reliance on domain expertise and minimizing user interference in the learning of constitutive laws. Neural-network-based methods have been successfully applied to learn various constitutive relationships, such as elasticity~\cite{ le2015computational, as2022mechanics}, viscoelasticity~\cite{chen2021recurrent} and plasticity~\cite{ghaboussi1991knowledge,bessa2017framework, mozaffar2019deep, li2022plasticitynet}. Additionally, a recent research line is to discover interpretable constitutive relations~\cite{fuhg2024extreme,kissas2024language}. However, ML methods typically depend on stress data for training, making them challenging to apply to experimental data where stress is not directly measurable. 
To address this challenge, a series of works have been developed to learn constitutive relations without using direct stress data, such as the different variants of EUCLID~\cite{flaschel2021unsupervised, flaschel2022discovering, joshi2022bayesian, thakolkaran2022nn}, and the work independently developed in~\cite{wang2021inference}. Here, we refer to these methods as unsupervised learning approaches since they do not rely on direct supervision from training labels, which, in this context, correspond to stress data.
In addition, we note an increasing interest in computer graphics for learning the constitutive relation to simulate or predict the motions of deformable objectives~\cite{du2023deep}. An exemplary work is the NCLaw~\cite{ma2023learning} that trains an MLP~\cite{bishop2006pattern} to approximate the constitutive relation. These unsupervised methods bypass the need for stress data by training the machine learning model to satisfy the linear momentum equation for a given displacement field.


\begin{table}[!h] 
  \centering
  \caption{Comparison between existing exemplary \emph{unsupervised} and proposed ML approaches for learning material models considering three desired properties. 
  } 
  \label{table:methods_comparison}
  \resizebox{1.0\textwidth}{!}{%
  \begin{tblr}{colspec={Q[m,c,6cm] Q[m,c,4cm]  Q[m,c,7cm] Q[m,c,3cm] },row{1}={3.5ex}, row{2-5}={5ex}, row{6}={9ex}, row{7}={5ex}}
    \toprule[1.5pt]
    Methods & No need for stress data for training  & No need for boundary forces for training & Enforce physics-consistency  \\ 
    \midrule
    EUCLID variants~\cite{flaschel2021unsupervised,flaschel2022discovering,joshi2022bayesian,thakolkaran2022nn}
    & Yes & No & Yes\\
    \hline
    NCLaw~\cite{ma2023learning}
    & Yes & Yes & No \\
    \hline
    Our approach & Yes & Yes & Yes\\
    \bottomrule[1.5pt]
    \end{tblr}
    }
\end{table}

Despite their great potential, unsupervised ML approaches also have limitations, as summarized in Table~\ref{table:methods_comparison}. Some methods depend on prior knowledge of the constitutive relation to design a library of candidate constitutive relations~\cite{flaschel2021unsupervised, flaschel2022discovering, joshi2022bayesian, wang2021inference}. This reliance on a predefined candidate library restricts the ML model to approximating only a limited set of possible constitutive relations.  Consequently, if the actual material behavior deviates significantly from this pre-assumed set of possible constitutive relations, the model may fail to accurately capture the true constitutive relation. 
NN-EUCLID~\cite{thakolkaran2022nn} overcomes this limitation by using an input convex neural network (ICNN)~\cite{amos2017input}, specifically designed to approximate convex functions and previously employed in a supervised context by~\citet{as2022mechanics}, as the surrogate constitutive model. 
While these methods circumvent the need for full-field stress data, they still rely on full-field displacement data along with information on boundaries (\textit{e.g.}, the reaction forces) to avoid learning trivial solutions for the constitutive relation (\textit{e.g.}, always outputting zero stress for any input strain in cases without external body force density)~\cite{flaschel2021unsupervised, flaschel2022discovering, joshi2022bayesian, wang2021inference,thakolkaran2022nn}~\footnote{Strictly speaking, Bayesian-EUCLID~\cite{joshi2022bayesian} is different from other EUCLID variants~\cite{flaschel2021unsupervised, flaschel2022discovering, wang2021inference,thakolkaran2022nn} because it uses dynamic data. However, \citet{joshi2022bayesian} consistently uses the boundary force for training, which, as demonstrated in this work, is unnecessary in the dynamic formulation.}. These two complementary data sources are not readily available in some situations, as for instance in \emph{in-situ} experiments on living biological materials. 
In contrast, NCLaw~\cite{ma2023learning}, which was developed primarily for computer graphics applications, bypasses the need for preparing a candidate constitutive relation library by using a standard MLP as the surrogate constitutive model. 
More importantly, it only uses temporal displacements for training.
However, this approach can result in learned constitutive relations that may look realistic but violate physical constraints. For instance, the learned energy density might not be convex with respect to strain invariants in hyperelasticity. 
To summarize, inferring a \emph{physics-consistent} material model becomes challenging when there is no prior knowledge about the constitutive relation and no access to complementary data such as stresses or boundary forces.


In this work, we aim to integrate the advantages of both EUCLID and NCLaw (see the last entry in Table~\ref{table:methods_comparison}) to develop a framework for learning \emph{physics-consistent} constitutive relations only using observable displacements. This is accomplished through two key designs. First, we consider a dynamic setting; specifically, the nodal force equilibrium condition for the \emph{internal domain} as the training objective. With this objective function, the proposed framework only requires using recorded temporal evolution of displacements to learn the constitutive relation, without any information about stress or boundary forces. Second, we use ICNN as the surrogate model, allowing for learning constitutive relations without prior knowledge of the material properties. This ensures the learned constitutive relation is \emph{physics-consistent}. 
To enforce the requirement that certain ICNN parameters remain non-negative,  projected gradient descent~\cite{beck2017first} is used as the training strategy. This method efficiently backpropagates gradients under the non-negativity constraint, enabling the framework to effectively learn various constitutive relations solely using strain invariants (without preprocessing input features) as the input feature.
Our approach is trained end-to-end, generalizes to different constitutive relations, and eliminates the reliance on prior knowledge of materials. Through extensive numerical simulations, we demonstrate the model's capability to uncover accurate strain-stress relations for various materials. Moreover, the learned constitutive relation converges to the ground-truth constitutive relation as we refine the data resolution used for training.

\section{Methodology: \underline{u}nsupervised \underline{l}earning of hyper\underline{e}lasticity in \underline{d}ynamics (uLED)}
\label{sec:method}

\begin{figure}[!ht]
    \centering
    \includegraphics[width=1\textwidth]{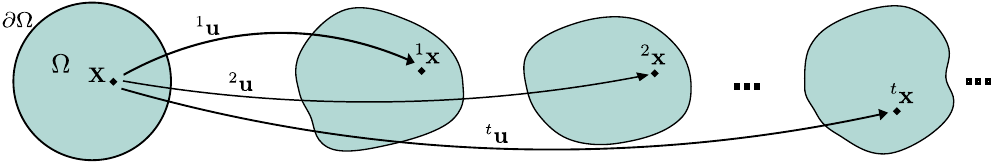}
    \caption{{Illustration of the reference configuration $\domain$ and  the deformation over time.} The displacement $\superscriptT{t}\vu$ of an arbitrary material point at each time $t$ is the difference between  its position $\superscriptT{t}\vx$ at time $t$ and its original position $\mX$ in the reference configuration $\domain$, \textit{i.e.}, $\superscriptT{t}\vu = \superscriptT{t}\vx - \mX$.}
    \label{fig:displacement}
\end{figure}

\subsection{Considered mechanical system}
\label{sec:problem_statement}
We consider a deformable solid with its reference configuration $\domain$ in $\mathbb{R}^d$ of dimension $d$, undergoing dynamic deformation over time and occupying the domain $\superscriptT{t}\domain$ at time $t$. The solid is subject to applied displacements on the Dirichlet boundary $\partial \domain_{D}$ and applied tractions on the Neumann boundary $\partial \domain_{T}$.
An arbitrary material point located at position $\mX$ in the reference domain $\domain$ is positioned at $\superscriptT{t}\vx$ in the current domain $\superscriptT{t}\domain$. The displacement of this material point at time $t$ is given by $\superscriptT{t}\vu$, where $\superscriptT{t}\vu =  \superscriptT{t}\vx -  \mX$ (see Fig.~\ref{fig:displacement}). The deformation gradient at time $t$ is defined as $\superscriptT{t}\mF =  \mI + \nabla_\mX  \superscriptT{t}\vu \in \mathbb{R}^{d\times d}$, or, in index notation, $\superscriptT{t}F_{ij} = \partial \superscriptT{t}u_{i}/\partial X_j \equiv \superscriptT{t}u_{i,j}$ ($i,j = 1, \ldots, d$). 
The stress in the solid is described by the first Piola-Kirchhoff stress tensor $\superscriptT{t}\mP = \cP(\superscriptT{t}\mF) \in \mathbb{R}^{d\times d}$, where $\cP$ represents the constitutive law relating the stress to the material's deformation. The solid must satisfy the conservation of linear momentum at time $t$, which, for the reference configuration, is given by:
\begin{equation}
\label{eq:governing_equ_strong_form}
    \forall i = 1, \ldots, d: \quad \rho \superscriptT{t}{}{\ddot{u}{}_i} - \superscriptT{t}P_{ij, j} - \superscriptT{t}B_i = 0 ~ \text{ in } \Omega ~,
\end{equation}
where $\superscriptT{t}B_i$ is the $i$-th dimension of the body force density $\superscriptT{t}\mB \in \mathbb{R}^{d}$ at time $t$. In addition, the stresses must  balance with the applied tractions on $\partial \domain_{T}$.
Unlike previous methods~\cite{thakolkaran2022nn, flaschel2021unsupervised, wang2021inference} that consider static equilibrium and require knowledge of boundary forces to learn the constitutive relation, we consider the system under  dynamic deformation (Eq.~\ref{eq:governing_equ_strong_form}). This approach  enables us to learn the constitutive relation without any information about boundary forces, as we will demonstrate.


\subsection{Objective and proposed method}
\label{sec:concept_of_proposed_method}
The objective of our proposed method is to infer the constitutive law that relates material stress to strain, \textit{i.e.}, $\cP(\mF)$, solely based on observations of the temporal evolution of the displacement field $\superscriptT{t}\vu$. Concretely, we aim to construct a neural-network-based surrogate model $\cPapprox(\mF; \theta)$ to approximate the constitutive material law, \textit{i.e.}, $\cPapprox(\mF; \theta) \approx \cP(\mF)$, where $\cPapprox(\mF;\theta)$ is parameterized by $\theta$. For a given experimentally measured displacement field $\superscriptT{t}\vugiven(\mX)$, we aim for $\cPapprox(\mF;\theta)$ to satisfy the conservation of linear momentum (Eq.~\ref{eq:governing_equ_strong_form}) for all times $t \leq T$. We denote the predicted first Piola-Kirchhoff stress by $\superscriptT{t}\widehat{\mP}$, $\superscriptT{t}\widehat{\mP} = \cPapprox(\superscriptT{t}\mF; \theta)\in \mathbb{R}^{d\times d}$.  In contrast to prior work, we limit the spatial region of interest $\subdomain$ to a subdomain of $\domain$, \textit{i.e.}, $\subdomain \subseteq \domain$, which leads to the following strong form:
\begin{equation}
    \forall i = 1, \ldots, d: \quad \rho\superscriptT{t}{}{\ddot{u}_i} -  \superscriptT{t}\widehat{P}_{ij,j} - \superscriptT{t}B_i = 0~ \text{ in } \subdomain  ~.
\label{eq:strong_opt_prob}
\end{equation}

Furthermore, since  we consider dynamic deformation where the term $\rho\superscriptT{t}{}{\ddot{u}_i}$ is nonzero, we do not need to know boundary forces to avoid learning the trivial solution\footnote{The trivial solution occurs  when the learned stress is a constant that only depends on the body force density, assuming the body force density is uniform throughout solid. In the case of static equilibrium (where $\rho\superscriptT{t}{}{\ddot{u}_i}$ is zero), boundary forces are  necessary to regularize the learned constitutive relation and  avoid such trivial solutions.}.
To solve this problem, we convert it into its weak form , which can then be solved as an optimization problem. The weak form of Eq.~\ref{eq:strong_opt_prob} is given by 
\begin{equation}
\label{eq:governing_equ_weak_form}
    \superscriptT{t}\functional_i(\theta; v_i) \equiv \int_{\subdomain}v_i \rho \superscriptT{t}{}{\ddot{{u}}_i} \dx + \int_{\subdomain}  v_{i,j}\superscriptT{t}\widehat{P}_{ij}  \dx - \int_{\subdomain}v_i \superscriptT{t}B_i \dx = 0~, 
\end{equation}
where $v_i$ is an arbitrary test function in the functional space $V = \{v_i \mid v_i = 0 \text{ on } \subdomainboundary \}$. It is important to note that $\superscriptT{t}\functional_i(\theta; v_i)$ depends on the learnable parameters $\theta$ because $\superscriptT{t}\widehat{P}_{ij}$ is the output of the neural-network-based surrogated constitutive model parameterized by $\theta$. Given that the displacements are known everywhere, including on $\subdomainboundary$, we are free to choose this $V$. This choice ensures that the unknown tractions from the boundary do not affect the conservation of linear momentum.
Thus, the optimization problem is formulated as follows: 
\begin{equation}
    \label{eq:optimization_problem}
    \text{Find $\theta$ such that } \superscriptT{t}\functional_i(\theta; v_i) = 0 \text{ for all } v_i \, .
\end{equation}


Finally, we note that the constitutive relation is learned to satisfy the principle of energy conservation through Eq.~\ref{eq:governing_equ_weak_form} without requiring stress data. Specifically, we calculate the kinetic energy $\int_{\subdomain}v_i \rho \superscriptT{t}{}{\ddot{{u}}_i} \dx$ based on nodal accelerations and use it to determine the expected internal strain energy $\int_{\subdomain}  v_{i,j}\superscriptT{t}{P}_{ij}  \dx$ (assuming the body force is zero). Consequently, with the surrogate model $\cPapprox(\mF; \theta)$ satisfying Eq.~\ref{eq:governing_equ_weak_form}, the predicted internal strain energy $\int_{\subdomain} v_{i,j}\superscriptT{t}\widehat{P}_{ij} \dx$  closely approximates its target value $\int_{\subdomain}  v_{i,j}\superscriptT{t}{P}_{ij}  \dx$. 

\begin{figure}[!ht]
\centering
\includegraphics[width=1\textwidth]{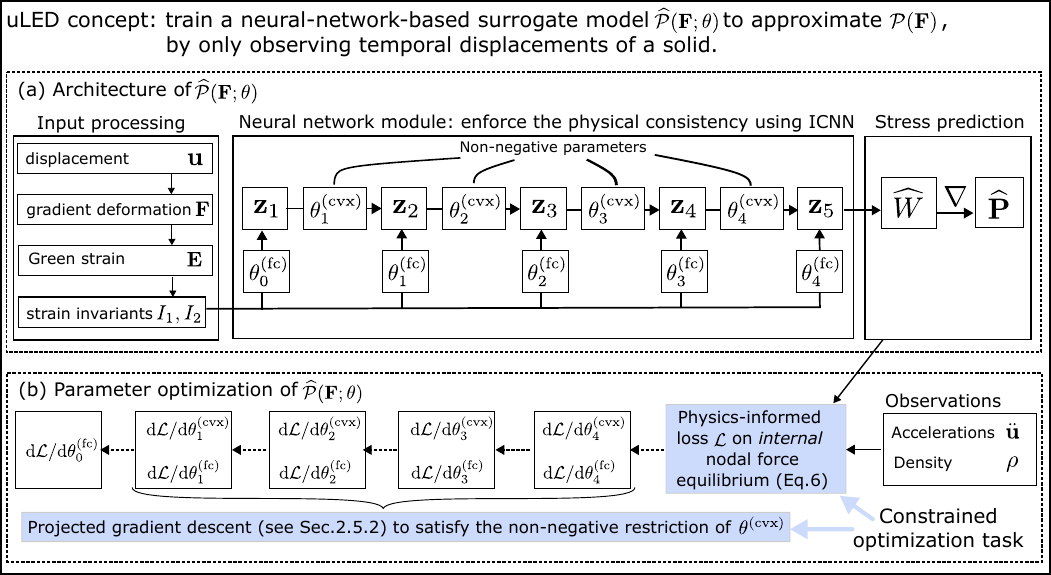}
\caption{{The concept and architecture of the proposed uLED method.} $\cPapprox(\mF; \theta)$ is the neural-network based surrogate model designed to approximate the constitutive material law, \textit{i.e.}, $\cPapprox(\mF; \theta) \approx \cP(\mF)$, where $\cPapprox(\mF;\theta)$ is parameterized by $\theta$. (a) The neural network architecture of the surrogate model $\cPapprox$. From the measured displacement field $\vu$, we compute the deformation gradient $\mF$, the Green-Lagrangian strain $\mE$, and strain invariants $I_1$ and $I_2$. The neural network (ICNN~\cite{amos2017input}) takes strain invariants as input and transforms them through  a series of latent representations $\vz_1, \vz_2, \ldots, \vz_5$ via multiple layers. The output at the last layer corresponds to the predicted energy density $\widehat{W}$, \textit{i.e.}, $\widehat{W}=\vz_5$. The predicted first Piola-Kirchhoff stress $\widehat{\mP}$ is computed by taking the gradient of the energy density. The learnable parameters $\theta$ in $\cPapprox(\mF; \theta)$ include the non-negative learnable layer-wise parameters 
$\theta^{\text{(cvx)}}$ and normal learnable parameters $\theta^{\text{(fc)}}$ without the non-negative restriction. (b) Parameter optimization of the surrogate constitutive model $\cPapprox(\mF; \theta)$. The learning objective (Eq.~\ref{eq:objective_function}) is formulated as a constrained optimization task and projected gradient descent is adopted to efficiently optimization parameters under the non-negative constraint.}
\label{fig:nn_architecture}
\end{figure}

\subsection{Physics-consistency requirements for the learned constitutive relation}
\label{sec:physics-consistency-requirement}
We aim for the surrogate model $\cPapprox(\mF; \theta)$ to satisfy fundamental physical constraints about the constitutive relation. Unfortunately, $\cPapprox(\mF; \theta)$ may violate fundamental physical constraints because it only minimizes Eq.~\ref{eq:optimization_problem}. Possible causes include imperfect training data such as noise or discretization error in the observed displacements. Following~\cite{as2022mechanics,thakolkaran2022nn}, we enforce $\cPapprox(\mF; \theta)$ satisfying the following three physical constraints:
\begin{enumerate}
    \item Objectivity. The energy density is rotation-invariant to the deformation gradient and the stress is rotation-equivalent to the deformation gradient.
    \item Triviality. The stress is zero when no deformation happens.
    \item Stability. Small loads do not lead to arbitrary deformations.
\end{enumerate}

\subsection{Neural network design for $\cPapprox(\mF; \theta)$}
\label{sec:nn_design}
Taking into account the three constraints outlined in Sec.~\ref{sec:physics-consistency-requirement}, we design uLED to learn the parameterized constitutive relation $\cPapprox(\mF; \theta)$  in three steps (see Fig.~\ref{fig:nn_architecture}): 
\begin{enumerate}
    \item Input processing: Transform the measured displacements into strain invariants
    \item Neural network module: Approximate the energy density function using the input convex neural network
    \item Output prediction: Extract the predicted energy density and compute the first Piola-Kirchhoff stress.
\end{enumerate}

\textit{Input processing.}
As mentioned in Sec.~\ref{sec:concept_of_proposed_method}, the proposed method only relies on the recorded temporal evolution of nodal displacements at the given data resolution. The velocity and the acceleration of the displacements are computed from the displacements at consecutive time steps. Starting from the nodal displacements, we can compute the deformation gradient by its definition (see Sec.~\ref{sec:problem_statement}). While it is generally convenient to use the Green–Lagrange strain tensor $\mE = (\mF^T\mF - \mI)/2$ or equivalently the right Cauchy–Green deformation tensor $\mC = \mF^T\mF$ to  represent deformation without rigid body movement, we use strain invariants as the input for the neural network. In the 2D case, the strain invariants are:  $ I_1 = \tr(\mC), I_2 = \frac{1}{2} (\tr(\mC)^2 - \tr(\mC^2))$. The advantage of using strain invariants is that they ensure the derived constitutive relation possesses  desired properties~\cite{satorras2021n}, such as  rotation-invariant predicted energy density   and  rotation-equivalent predicted stress  with the input deformation gradient. 
Particularly, our inputs to the neural network are solely strain invariants, without any additional transformation or feature engineering. This greatly reduces the need for domain knowledge of different materials. 


\textit{Neural network module.} Following~\cite{as2022mechanics,thakolkaran2022nn}, we enforce the predicted energy density $\widehat{W}$ to be convex in the Green-Langrange strain tensor $\mE$ in order to satisfy the stability constraint. To this end,
we use the Input Convex Neural Network (ICNN)~\cite{amos2017input} to approximate the ground-truth energy density as a function of strain invariants. ICNN takes the strain invariants as input and outputs the energy density, \textit{i.e.}, $\widehat{W} = \text{ICNN}(I_1, I_2; \theta)$, where $\theta$ are the learnable parameters. The reason for choosing ICNN over other types  of neural networks is that ICNN guarantees the output is a convex function of the input. Convexity in the strain invariants ensures local material stability~\cite{as2022mechanics, thakolkaran2022nn}. However, achieving  the desired convexity with ICNN comes with the restriction that part of its parameters ($\theta^{\text{(cvx)}}$ in the blue box of Fig.~\ref{fig:nn_architecture}) must be non-negative. This restriction makes training ICNN non-trivial. 
Specifically, we use the projected gradient descent~\cite{beck2017first} to enable the gradients efficiently backpropagated under the non-negative constraint. We introduce the optimization strategy in Sec.~\ref{sec:training_strategy} and the detailed implementation of ICNN is explained in~\ref{sec:SI_ICNN_configuration}.

\textit{Stress prediction.}
For hyperelastic materials, the first Piola-Kirchhoff stress $\mP$ can be computed as the gradient of the elastic energy density $W$, given by $\mP = \frac{\partial W}{\partial \mF}$, where $W = \mathcal{W}(\mF)$ depends on the deformation gradient $\mF$.  Equivalently, $\mP = \mF\mS$, where $\mS = \frac{\partial W}{\partial \mE}$ is the second Piola-Kirchhoff stress associated with the Green–Lagrange strain tensor $\mE$. The predicted energy density and the predicted first Piola-Kirchhoff stress should be zero in the absence of deformation, \textit{i.e.}, $W =0$ and $\mP = \mathbf{0}$ when $\mF = \mI$ or equivalently $\mE = \mathbf{0}$. To enforce this property, we use the predicted energy density and stress in the undeformed state as a reference to correct the predictions for any other state~\cite{as2022mechanics, thakolkaran2022nn}. Specifically, we denote the deformation gradient with no deformation by $\undeformedF$ and the Green strain with no deformation by $\undeformedE$ (\textit{i.e.}, $\undeformedF = \mI$, $\undeformedE = \mathbf{0}$), and the strain invariants of $\undeformedE$ by $I_1^{\undeformedSymbol}, I_2^{\undeformedSymbol}$  (\textit{i.e.}, $I_1^{\undeformedSymbol} = 0$ and $I_2^{\undeformedSymbol} = 0$). The predicted energy density for $\undeformedE$ is denoted by $\widehat{W}^{\undeformedSymbol}$, where $\widehat{W}^{\undeformedSymbol} = \text{ICNN}(I_1^{\undeformedSymbol}, I_2^{\undeformedSymbol}; \theta)$. The corresponding predicted first Piola-Kirchhoff stress $\widehat{\mP}^{\undeformedSymbol}$ for $\undeformedE$ is computed by $\widehat{\mP}^{\undeformedSymbol} = \partial \widehat{W}^{\undeformedSymbol} / \partial \undeformedF$. For any general state characterized by the principal invariants $I_1$ and $I_2$, the predicted energy density $\widehat{W}$ is computed by $\widehat{W} = \text{ICNN}(I_1, I_2; \theta) - \widehat{W}^{\undeformedSymbol} - \mathrm{Trace}(\mE^\mathsf{T} \widehat{\mP}^{\undeformedSymbol})$. The corresponding predicted first Piola-Kirchhoff stress $\widehat{\mP}$ is computed by $\widehat{\mP} = \partial \widehat{W} / \partial \mF - \mF \widehat{\mP}^{\undeformedSymbol}$. This ensures that the predicted energy density and the first Piola-Kirchhoff stress are zero in the absence of deformation.

\subsection{Learning the optimal model}
\label{sec:learn_optimal_theta}
After defining the surrogate model for the constitutive relation in Sec.~\ref{sec:nn_design}, we need to find the optimal values for parameters $\theta$ in $\cPapprox(\mF; \theta)$ to solve the optimization problem defined in Eq.~\ref{eq:optimization_problem}. This problem presents two major challenges: 1) how to transform  the optimization problem in Eq.~\ref{eq:optimization_problem} into a numerically computable objective function and 2) how to ensure  the non-negative requirements for parameters $\theta^{\text{(cvx)}}$. In the following, we outline our approach to address these challenges.

\subsubsection{Training objective: constrained optimization task}
\label{sec:training_objective}
To numerically solve the optimization problem defined in Eq.~\ref{eq:optimization_problem}, we use the finite element method (FEM) to approximate $\functional(\theta; v_i)$ as defined in Eq.~\ref{eq:governing_equ_weak_form} as $\functionalApprox(\theta; v_i)$. Suppose the region of interest $\subdomain$ is discretized by a set of elements $\mathcal{C}_e$ with $n_e = |\mathcal{C}_n|$ elements and a set of nodes $\mathcal{C}_n$ with $n_n = |\mathcal{C}_n|$ \emph{internal} nodes. We denote the shape function associated with node $a\in \mathcal{C}_n$ by $N^a$ and the support elements on which $N^a$ is non-zero by $\text{supp}(N^a)$, \textit{i.e.}, $\text{supp}(N^a) = \{\text{el} \in \mathcal{C}_e :\mathrm{for~any~} \mX \in \text{el},  N^a(\mX) \neq 0\}$.  
Thus, we get the approximated functional $\functionalApprox(\theta; v_i)$ to  approximate $\functional(\theta; v_i)$, \textit{i.e.}, $\superscriptT{t}\functionalApprox(\theta; v_i) \approx \superscriptT{t}\functional(\theta; v_i)$, as follows: 
\begin{equation}
\label{eq:momentum_balance_every_node_internal}
 \begin{split}
     &\forall i=1, 2, \ldots, d:\quad \quad \superscriptT{t}\functionalApprox_i(\theta) \equiv \sum_{a=1}^{n_n} v_i^a \superscriptT{t}f_i^{a}(\theta)\, , \\
     &\superscriptT{t}f_i^{a}(\theta) = \sum_{\text{el} \in \text{supp}(N^a)} \left[\int_{\text{el}}N^a \rho N^b \dx  \superscriptT{t}{}{\ddot{u}{}^b_i} + \int_{\text{el}} N^a_{,j}  \superscriptT{t}\widehat{P}_{ij} \dx - \int_{\text{el}}N^a \superscriptT{t}B_i \dx\right]
 \end{split}
 \end{equation}
where $\prescript{t}{}f_i^{a}(\theta)$ is the $i$-th component of the nodal force at node $a$ at time $t$, with $\widehat{P}_{ij}$ being  the components of the predicted stress from the surrogate constitutive model $\cPapprox(\mF, \theta)$. The computation of integrals in Eq.~\ref{eq:momentum_balance_every_node_internal} is discussed in~\ref{sec:background}. 
Since the test function $v_i^a$ can have arbitrary values for internal nodes, $\superscriptT{t}f_i^{a}(\theta)$ in Eq.~\ref{eq:momentum_balance_every_node_internal} must be zero for these internal nodes at all times $t$, \textit{i.e.}, $\forall a \notin \partial\subdomain, \forall i\in d, \forall t:\, \superscriptT{t}f_i^{a} = 0 $. Therefore, we optimize the parameters $\theta$ to minimize the absolute value of $\superscriptT{t}f_i^{a}$ for internal nodes at all times $t$:
\begin{equation}
\label{eq:objective_function}
\begin{split}
    \argmin_{\theta = \{\theta^{\text{(cvx)}}, \theta^{\text{(fc)}}\}} \mathcal{L}(\theta) = \frac{1}{n_n}\frac{1}{d}\frac{1}{T}\sum_{a=1}^{n_n}\sum_{i=1}^d\sum_{t=1}^T |\superscriptT{t}f_i^{a}(\theta)| \; , ~~ s.t.~~  \theta^{\text{(cvx)}} \geq 0\; .
\end{split}
\end{equation}
where $\superscriptT{t}f_i^{a}(\theta)$ is defined in Eq.~\ref{eq:momentum_balance_every_node_internal}. As indicated by the objective loss function (Eq.~\ref{eq:objective_function}), the only required data for training the neural network are the nodal displacements $\{\superscriptT{t}\vu^{a}\}$ and accelerations $\{\superscriptT{t}\ddot{\vu}^{a}\}$ of the internal nodes $\{a \mid a \in \subdomain, a \notin \subdomainboundary\}$ at several time steps $t=1, 2, \ldots, T$. This approach circumvents the need for any information about the stress data or boundary forces. 
Notably, we directly minimize the loss based on the governing equation instead of computing the simulated displacements and minimizing the discrepancy with the ground-truth displacements, as commonly done. The advantage of our approach is that we do not need to solve the governing equation using a FEM solver, which saves computational cost and avoids difficulties in backpropagating gradients through the FEM solver.
Additionally, the region of interest $\subdomain$ can be a subset of $\domain$, indicating that full-field displacement observation is not necessary, and we can select a part of the material to infer the constitutive law. This property is particularly  useful in cases where we can only measure the dynamics of a part of a material.

\subsubsection{Training strategy to satisfy the non-negative restriction}
\label{sec:training_strategy}
We use gradient-based methods to efficiently find the optimal values for $\theta$. However, compared to training a standard neural network, we have special constraints on some parameters (the non-negative parameters $\theta^{\text{(cvx)}}$) in ICNN. 
Two possible strategies exist to satisfy this non-negativity constraint. The first strategy is reparameterization which uses an additional transformation, such as the Softplus~\cite{nair2010rectified} used in~\cite{as2022mechanics, thakolkaran2022nn} and maps $\theta^{\text{(cvx)}}$ to non-negative values. With the transformed variable as the new optimization variable, this approach ensures the non-negativity constraint is always met for any value of $\theta^{\text{(cvx)}}$. 
The second approach is the projected gradient descent method~\cite{beck2017first} which was used by~\cite{klein2022polyconvex, fuhg2024extreme} and is adopted here based on empirical results showing that it outperforms the reparameterization approach (see Sec.~\ref{sec:SI_training_strategy}). Specifically, we perform two additional steps compared to the standard gradient descent method to train an ICNN to satisfy these constraints on the non-negative weights. 
First, when randomly initializing parameters before training, we enforce $\theta^{\text{(cvx)}}$ is initialized with non-negative weights. 
Second, after each gradient descent step, we project the weights $\theta^{\text{(cvx)}}$ onto the non-negative space. Specifically, if a parameter value is negative, we set  it to 0; if the value is non-negative, we leave it unchanged.
These two steps ensure that the non-negative weights $\theta^{\text{(cvx)}}$ in ICNN, thereby guaranteeing that the predicted energy density is a convex function of the input. The pseudo-code for the training procedure is summarized in~\ref{sec:SI_training_details}.

\subsection{Hyperparameter Settings}
We maintain consistent hyperparameter settings across different experiments. This includes using a uniform configuration for the ICNN in all experiments, including the number of layers, layer width, and the activation function. This standardized approach helps to assess the model's robustness across different testing scenarios.
Further details on the ICNN configuration are provided in Sec.~\ref{sec:SI_ICNN_configuration}. 
For training  hyperparameters such as the learning rate, we adjust only the number  of training epochs to ensure convergence across  different experiments. Comprehensive  information about these training hyperparameters is available  in~\ref{sec:SI_training_details}.

\section{Validation and evaluation of the proposed framework}
\label{sec:validation}

\begin{figure}[!ht]
\centering
\includegraphics[width=1.0\textwidth]{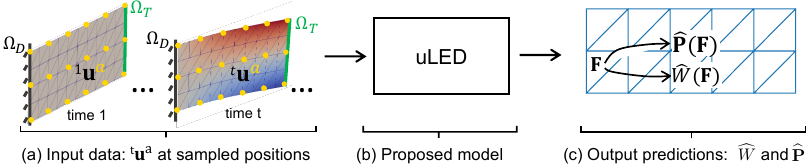}
\caption{{Illustration of the experimental configuration.} (a) We simulate the transient motion of a 2D plate that is fixed at its left boundary and loaded with a temporally evolving traction on its right boundary. We record the displacements $\superscriptT{t}\vu^{a}$ of sampled material points (indicated by the yellow dots) from the FEM simulation. These sampled displacements $\superscriptT{t}\vu^{a}$ are used as the input for uLED. (b) The proposed uLED method is trained using   the sampled displacements $\superscriptT{t}\vu^{a}$. (c) The model predicts the energy density value and the first Piola-Kirchhoff stress tensor for an arbitrary element given the deformation gradient.}
\label{fig:experiment_illustration}
\end{figure}
To showcase uLED's performance and versatility, we evaluate it with synthetic data from numerical simulations. We generate simulation results for a given material model, which constitutes the ground-truth constitutive relation. Then, we train uLED from the generated data and validate whether the learned constitutive relation approximates the ground-truth constitutive relations with sufficient accuracy.

To this end, we use FEM to generate a temporally evolving displacement field of a 2D plate as shown in Fig.~\ref{fig:experiment_illustration}. We fix the left boundary of the plate and apply transient tractions on the right boundary. We do not consider body forces in the simulation, \textit{i.e.}, $\superscriptT{t}\mB \equiv \mathbf{0}$. 
Without loss of generality, we use dimensionless units for all simulations. Details of data generation are outlined in~\ref{sec:SI_data_generation}. The FEM simulation represents the underlying physics process, from which we sample a set of material points and record nodal displacements at different time steps. Nodal velocities and accelerations can be computed from the displacements. The sampled nodal displacements and accelerations are then used to train uLED.
For each constitutive relation, the sampled nodal displacements and accelerations are collected from 1,500 simulation time steps. To reduce training time, we construct a dataset by selecting every 7-th time step, resulting in 214 time steps. The first 80\% of these 214 time steps are used for training the model, and the remaining 20\% are used for validation. For each material, we conduct three independent experiments, randomly initializing the learnable parameters in uLED and training the model with the same training data. In each experiment, the validation data is used to select the best-trained model from the entire training period. Details to construct the uLED input from FEM simulation are presented in \ref{sec:construct_input_from_FEM}.


In real-world applications, the recorded data's resolution is determined by the measurement equipment, while the true physical process occurs typically at smaller scales. We reproduce this situation with our synthetic data by carefully choosing and distinguishing between the data resolution used during the generation and as the uLED input. During data generation using FEM simulations, we aim to accurately represent the relevant underlying physics and, hence, use a high resolution, referred to as $\frac{1}{h_{gen}}$ ($h_{gen}$ is the element size of the mesh). We use $h_{gen}=0.001$ consistently throughout this paper. To imitate the downsampling that generally occurs through the resolution of the measurement equipment, we use a lower resolution $\frac{1}{h}$ as input for uLED. Specifically, we set $\frac{1}{h} = \frac{1}{2 h_{gen}}$ throughout this paper, except in Sec.~\ref{sec:exp1_discretization}, where we explore the influence of data resolution on uLED. 


\begin{figure}[!h]
\centering
\includegraphics[width=1.0\textwidth]{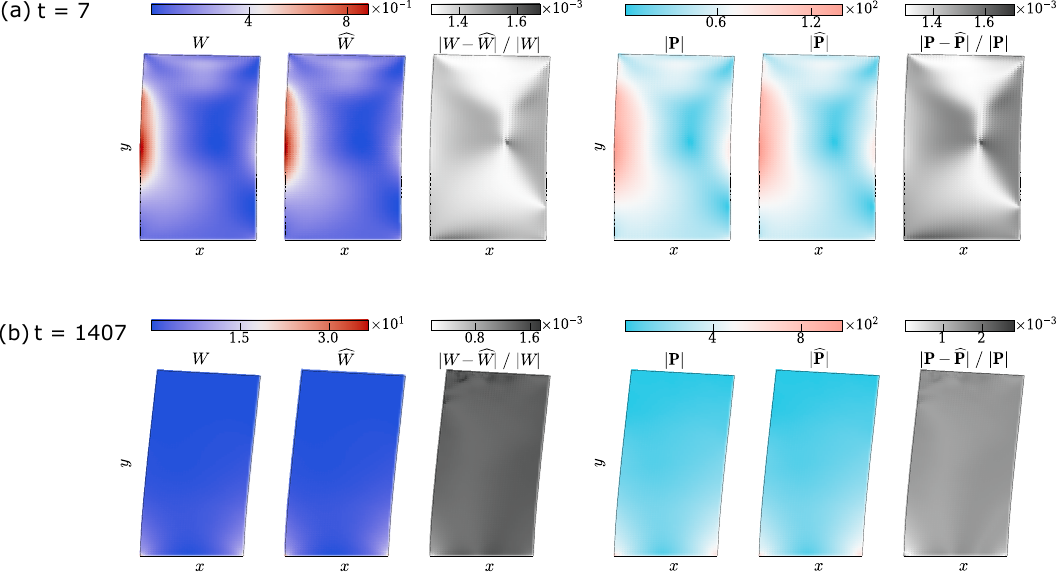}
\caption{{Comparison between the predicted and ground-truth energy density and stress fields for a Neo-Hookean material.} Data is shown for two arbitrarily chosen time steps (a) $t=7$, and (b) $t=1407$. On the left side, the predicted energy density ($\widehat{W}$) is compared to the ground-truth value ($W$) via their normalized difference. On the right side, the magnitude of the first Piola-Kirchhoff stress fields ($|\widehat{\mP}|$) is compared to its ground-truth values ($|\mP|$). The predicted values shown are the mean of three independent experiments.}
\label{fig:exp0_W_P_field_same}
\end{figure}

\subsection{Showcase example: Learning the Neo-Hookean constitutive relation}
\label{sec:NH_showcase}

In the first experiment, we use the Neo-Hookean material (see first entry in Table~\ref{table:hyperelatic_models} for its definition) as an example to thoroughly examine the effectiveness of uLED in learning the constitutive relation before we generalize the results to a set of other constitutive laws in subsequent sections. We directly use strain invariants $I_1$ and $I_2$ as the inputs to ICNN and evaluate the learned constitutive relation through two investigations. First, we randomly select two deformation states used in training and compare the predicted energy density and stress fields by the trained uLED model to the ground-truth energy density and stress fields computed with the Neo-Hookean constitutive relation. The difference between the predicted and ground-truth values is three orders of magnitude smaller than the scale of the ground-truth values (see Fig.~\ref{fig:exp0_W_P_field_same}). This indicates that the uLED model after training can effectively approximate the ground-truth constitutive relation, successfully learning both the energy density and stress with high accuracy.

\begin{figure}[!ht]
\centering
\includegraphics[width=1.0\textwidth]{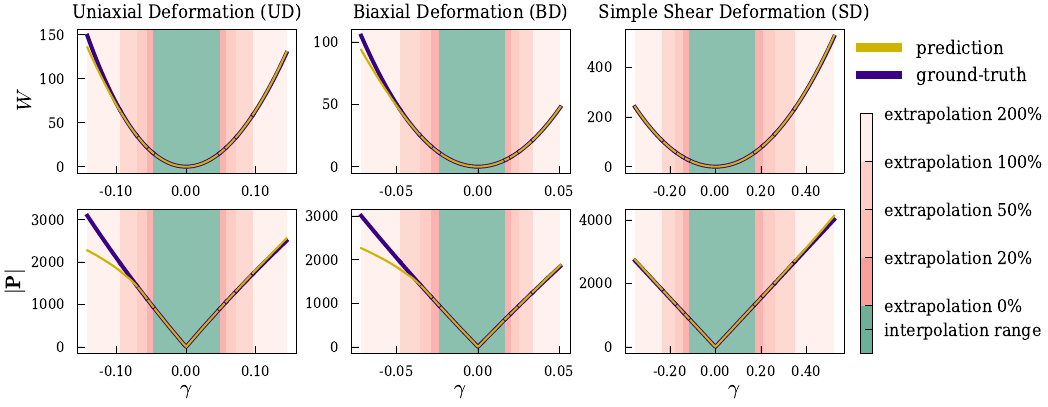}
\caption{{Validation of the predicted energy density and stress of the Neo-Hookean material.} We compare curves of the prediction to the ground-truth of the energy density (top) and the stress (bottom) for different deformation states (see Eq.~\ref{eq:F_path}) at a single material point. The predictions shown are the mean of three independent experiments. }
\label{fig:exp1_NH}
\end{figure}

As second evaluation, similar to~\cite{thakolkaran2022nn}, we compare the predicted and ground-truth energy density and stress values for multiple deformation states at \emph{a single material point}. We construct three  types of deformation: uniaxial deformation (UD), biaxial deformation (BD), and simple shear deformation (SD), defined as follows:
\begin{equation}
\label{eq:F_path}
\mF^{\text{UD}} = \begin{bmatrix}
    1 + \gamma & 0 \\
    0 & 1 \\
\end{bmatrix},\,\,\,\,\,\,
\mF^{\text{BD}} = \begin{bmatrix}
    1 + \gamma & 0 \\
    0 & 1 + \gamma\\
\end{bmatrix},\,\,\,\,\,\,
\mF^{\text{SD}} = \begin{bmatrix}
    1 & \gamma \\
    0 & 1\\
\end{bmatrix}
\end{equation}
where $\gamma$ is varied to generate multiple deformation states. The range of  $\gamma$ values is determined from the training data. Specifically, we compute the maximum and minimum  values of $\gamma$ that appeared in the training data to evaluate the interpolation ability. Additionally, we extend the range by up to 200\%,  to generate deformation gradients that are not present in the training data, allowing us to evaluate the extrapolation ability of uLED. Results show that the curve of the learned energy density function and the stress function (see Fig.~\ref{fig:exp1_NH}) match the corresponding ground-truth curves closely in both the interpolation and relatively large extrapolation regions (\textit{i.e.}, 50\% away from the training data range).

\subsection{Transferability to unseen geometries}
\begin{figure}[!h]
\centering
\includegraphics[width=1.0\textwidth]{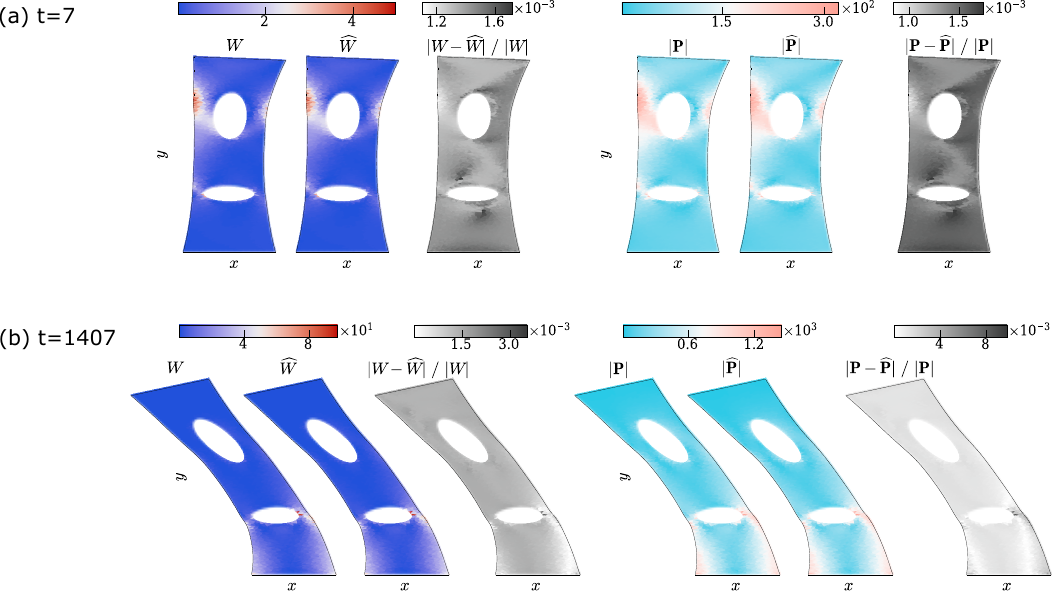}
\caption{{Results for transferability on the unseen material sample (i.e. geometry).} We train uLED on the plate sample and test the trained uLED on a new sample with holes. These two material samples used for training and testing are governed by the same Neo-Hookean law. We compare the predicted energy density field $\widehat{W}$ to the ground-truth $W$ and the predicted stress field magnitude $|\widehat{\mP}|$ to the ground-truth $|\mP|$ on this new sample  for two arbitrarily chosen time steps (a) t = 7, and (b) t = 1407. The predicted values shown are the mean of three independent experiments.}
\label{fig:exp0_W_P_field_different}
\end{figure}

One important characteristic of the learned material model is its transferability. This means that the uLED model trained on one sample of material should be applicable to other samples of the same material. For instance, if we have several samples of the same material, we can train uLED based on the observed dynamics of one sample. After this training, we expect that uLED can be directly applied to the other samples without additional training.

To demonstrate the transferability of uLED, we analyze the predicted energy density and stress on an \emph{unseen} sample with a different geometry from the sample used for training.  Specifically, we train uLED on a Neo-Hookean material sample with the plate geometry shown in Fig.~\ref{fig:experiment_illustration}~(a). After training, we apply the trained model to predict the energy density and the stress fields of a new sample governed by the same Neo-Hookean law, using the plate geometry shown in Fig.~\ref{fig:exp0_W_P_field_different}. Our results show that the prediction on this new sample closely matches the ground-truth,  with differences that are three orders of magnitude smaller. This experiment indicates that the constitutive relation learned by uLED is independent of the specific material sample used for training. Instead, uLED effectively learns the underlying constitutive law that governs the material behavior.

\subsection{Generalizability to learn constitutive relation for various materials}
\label{sec:exp2_various_materials}


Next, we investigate the generalizability of uLED across  different types  of materials. We consider various hyperelastic constitutive relations, as listed in \Cref{table:hyperelatic_models}, which cover a wide range of material behavior features. Importantly, we maintain  the same ICNN architecture (including the number of layers, layer widths, activation function, etc.) for all considered material models. Additionally, we highlight that we consistently use strain invariants $I_1$ and $I_2$ as the inputs to ICNN without any further feature engineering.

\begin{table}[!h] 
  \centering
  \caption{Hyperelatic constitutive models in 2D. $\lambda$ and $\mu$ are Lam\'e's first and second parameter, respectively. $K$ is the bulk modulus. $I_1$ and $I_2$ are the strain invariants. $J = \det(\mF)$ is the determinate of the deformation gradient. $\bar{I}_1 = J^{-1} I_1$ and $\bar{I}_2 = J^{-2} I_2$ are the equivalent invariants of the unimodular $\bar{\mC} = (\det(\mC))^{-1/2}\mC$.}
  \label{table:hyperelatic_models}
  \resizebox{1\textwidth}{!}{%
  \begin{tblr}{colspec={Q[m,c,6cm] Q[m,c,10cm] Q[m,c,3cm]},row{1}={3.5ex}, row{2-3}={4ex}, row{4}={6ex}, row{5}={4ex}, row{6}={7.5ex}, row{7}={4ex}, 
  cell{4}{3} = {font=\linespread{1.5}\selectfont},
  cell{6}{2} = {font=\linespread{1.5}\selectfont}
  }
    \toprule[1.5pt]
    Model & Energy density & Free parameters  \\
    \midrule
    Neo-Hookean (NH)~\cite{logg2012automated} & $W = \frac{\mu}{2} (I_1 - 2) - \mu \ln(J) + \frac{\lambda}{2} \ln(J)^2$ & N/A\\
    \hline
    St.~Venant–Kirchhoff (StVK)~\cite{logg2012automated} & $W = \frac{\lambda}{2} \tr(\mE)^2 + \mu \tr(\mE^2 )$ & N/A \\
    \hline
    Two parameter Mooney-Rivlin (MR)~\cite{bower2009applied} & $W = C_{10} (\bar{I}_1 - 2) + C_{01} (\bar{I}_2 - 1) + 0.5K(J - 1)^2$ &\makecell{$C_{10} = \frac{7}{16} \mu$ \\ $C_{01} = \frac{1}{16}\mu$}\\
    \hline
    Gent~\cite{gent1996new} & $W = - \frac{\mu}{2} (J_m \ln( 1 - ( \bar{I}_1 - 2) / J_m )  ) + 0.5 K (J - 1)^2$ & $J_m = 10$\\
    \hline
    Arruda-Boyce (AB)~\cite{bower2009applied} & $W = \mu ( \frac{1}{2} (\bar{I}_1 - 2) + \frac{1}{20N} ({\bar{I}{}}_1^2  - 4) + \frac{11}{1050 N^2} ({\bar{I}{}}_1^3  - 8) + \frac{19}{7000N^3} ({\bar{I}{}}_1^4  - 16) + \frac{519}{673750 N^4} ({\bar{I}{}}_1^5  - 32) ) + 0.5 K (J - 1)^2$ & $N = 10$\\
    \hline
    Fung~\cite{fung2013biomechanics} & $W = \frac{\mu}{2b} ( b (\bar{I}_1 - 2) + e^{b (\bar{I}_1 - 2)} - 1 ) + 0.5 K (J - 1)^2$ & $b = 1$\\
    \bottomrule[1.5pt]
    \end{tblr}
    }
\end{table}

\begin{figure}[!ht]
\centering
\includegraphics[width=1.0\textwidth]{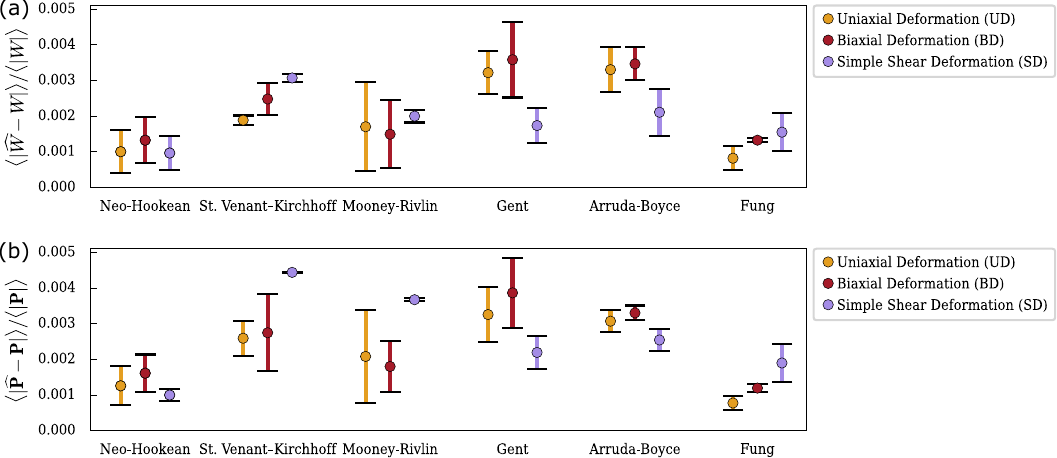}
\caption{Validation of the predicted constitutive relation for various materials. We compute the NMAE of (a) the predicted energy density and (b) the predicted stress compared to their ground-truth values for different deformation states at a single material point. We report the mean (indicated by the position of circles) and the standard deviation (indicated by the error bars) of the normalized MAE from three independent experiments.  
}
\label{fig:exp1_W_error_Fpath}
\end{figure}

We evaluate the performance of uLED for each constitutive relation by comparing the difference between the predicted and ground-truth energy density and stress values for multiple deformation states at a single material point (as described in detail in Sec.~\ref{sec:NH_showcase}). To measure uLED's performance, we use the normalized mean absolute error (NMAE) between the prediction and the ground-truth. For each deformation state, $\superscriptT{i}\mF$, we compute the predicted energy density, $\superscriptT{i}\widehat{W}$, and the ground-truth energy density, $\superscriptT{i}W$, corresponding to the deformation gradient (as given by Eq.~\ref{eq:F_path}). The NMAE is computed as the mean absolute error $\average{|\widehat{W} - W|}$, \textit{i.e.}, $\average{|\widehat{W} - W|} = \frac{1}{\Upsilon}\sum_{i=1}^{\Upsilon}|\superscriptT{i}\widehat{W} - \superscriptT{i}W|$, divided by the mean value of the ground-truth energy density $\average{|W|}$. Mathematically, this is expressed as: $\average{|W|} = \frac{1}{\Upsilon}\sum_{i=1}^{\Upsilon}|\superscriptT{i}W|$. The results indicate that the MAE between the predicted and ground-truth energy density values is three orders of magnitude smaller than the average ground-truth energy density values for all materials considered (see Fig.~\ref{fig:exp1_W_error_Fpath}~a). The same analysis for the predicted stress demonstrates similar findings, as shown in Fig.~\ref{fig:exp1_W_error_Fpath}~b.  Additional analyses of the curves of learned energy density function and stress function can be found in \ref{sec:SI_W_P_along_Fpaths}. 
Additionally, we observe a significant decrease in both training loss and validation loss during the training process, indicating that the proposed optimization strategy is effective in training uLED to learn various constitutive relations using only strain invariants as input.
Detailed training curves illustrating this process are presented in~\ref{sec:SI_training_curves}.




\section{Towards applicability to experimental data}
Experimental data is typically less  precise and more noisy compared to the finite element method (FEM) data that was artificially generated and utilized in Sec.~\ref{sec:validation} to evaluate the capabilities of uLED. Imperfections in training data can arise from several factors,  including coarse data resolution in the measurement of the displacement field, limited  observation windows, 
the noise introduced by the digitalization of instruments,
and  dissipative processes in the material due to internal friction. This section examines how these factors impact uLED's capacity to discover constitutive relations.

\subsection{Effect of data resolution}
\label{sec:exp1_discretization}

We begin  by investigating the effect of reduced data resolution on the performance of uLED. Specifically, we run a FEM simulation on a fine mesh ($h_{gen} = 0.001$) and train uLED using  different levels of coarsened data resolution $1/h$ where $h = 2h_{gen}, 3h_{gen}, \ldots, 10h_{gen}$. For this investigation, we use the Neo-Hookean constitutive law as an example,  though the results would be applicable to any of the other  constitutive laws considered (Table~\ref{table:hyperelatic_models}).

\begin{figure}[h!]
\centering
\includegraphics[width=1\textwidth]{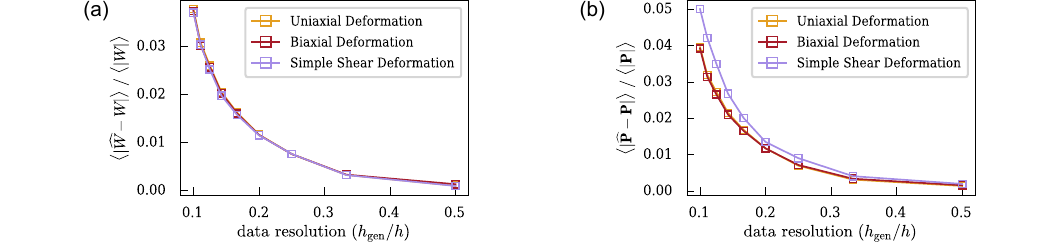}
\caption{{Convergence evaluation regarding data resolution with Neo-Hookean material.}
We compute the mean absolute error of the predicted (a) energy density and (b) stress for different deformation states at a single material point, normalized by the average value of the ground-truth value. The predicted values are the mean of three independent experiments.}
\label{fig:exp2_convergence}
\end{figure}

As expected, the data resolution has direct consequences on the quality of the learned energy density and stress functions. Similar to Sec.~\ref{sec:exp2_various_materials}, we compare the energy density and stress values predicted by the uLED model, trained with different data resolutions, against the ground-truth values for multiple deformation states at \emph{a single material point}. As shown in Fig.~\ref{fig:exp2_convergence}, the normalized mean absolute error of the predicted energy density and stress consistently decreases with higher data resolution $1/h$ for all considered deformation states. 
Additional analysis about the learned energy density and stress fields with different data resolutions can be found in \ref{sec:SI_convergence_data_resolution}.
These results illustrate that uLED learns the constitutive laws as well as possible given the quality of the available data -- even relatively low-resolution data may still provide good predictions of the energy density and stress. More importantly, our analysis demonstrates that the learned constitutive relation converges to the ground truth as we refine the data resolution used for training uLED. This suggests that uLED is potentially applicable to data with relatively low resolutions, as measured from physical experiments, and that the results can be improved by refining the data acquisition.

\subsection{Learning constitutive relation with a limited observation window}
\begin{figure}[!t]
\centering
\includegraphics[width=1.0\textwidth]{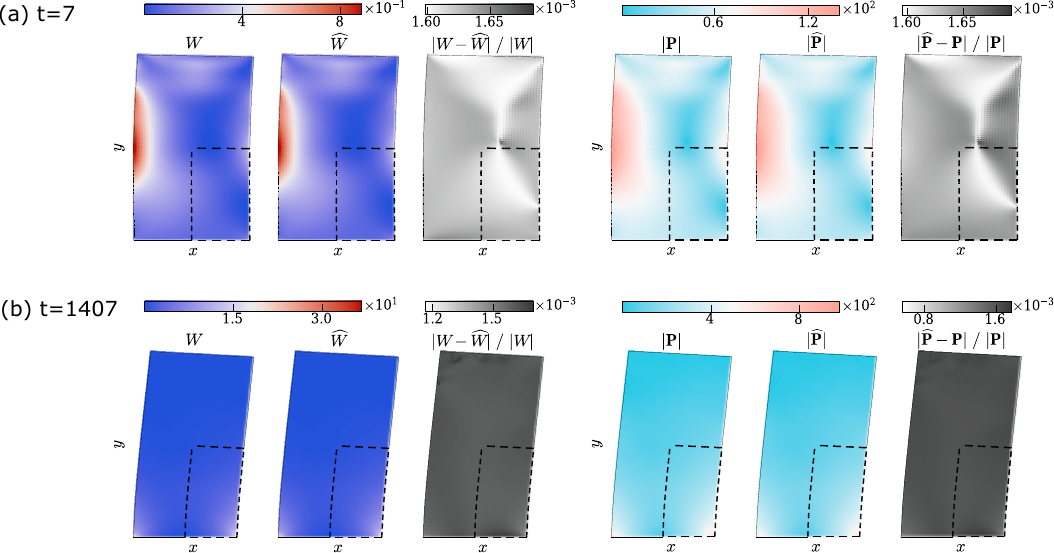}
\caption{
Results for validation with partial observation on the Neo-Hookean material case. The uLED model is trained using the displacement fields from the area marked with the dashed box only (excluding the border of the box). The model predicts the energy density for the full domain. The predicted values represent the mean of three independent experiments. For comparison purposes, the selected time steps (a) t=7, and (b) t = 1407 are identical to those in the reference case with training on the full domain (Fig.~\ref{fig:exp0_W_P_field_same}).}
\label{fig:exp5_partial_observation_fields}
\end{figure}

An important challenge for inference methods is that experimental data, particularly for in-situ observations, is often limited to a part of the (test) specimen. The formulation of uLED offers a significant advantage in that it does not require a full displacement field to learn a constitutive law. To evaluate this capability of uLED, we consider a scenario where the data is only available from a subdomain of the material. Specifically, we train uLED using the deformation data from only a quarter of the full domain (marked by the dashed box in Fig.~\ref{fig:exp5_partial_observation_fields}). For evaluation, we use the trained model to predict the energy density and the stress across the entire domain (see Fig.~\ref{fig:exp5_partial_observation_fields}). The results show that, compared to the reference case with training on the full domain,  the predicted energy density and stress exhibit a slightly larger error (see the normalized errors in Fig.~\ref{fig:exp0_W_P_field_same} and Fig.~\ref{fig:exp5_partial_observation_fields}, noting that the ground-truth energy density fields are identical). This outcome is expected because the training data is more limited in this scenario, making the model trained with partial observation more likely to make inaccurate predictions for the deformation gradients not covered by the training data. 
However, the difference between the predicted and ground-truth values for the energy density and stress fields is still three orders of magnitude smaller than the scale of the ground-truth values. 
These results indicate that uLED can function with partial observation data well and the accuracy improves with more comprehensive training data.

\subsection{Effect of noisy data}
\label{sec:exp3_noise}

The quality of the learned model is directly related to the quality of the data. Since the inference of material constitutive laws  is aimed at  applications involving  experimentally acquired data, which is unavoidably noisy, it is crucial to evaluate the effect of such imperfections on the learning capabilities. One specific example of noisy experimental data originates from DIC measurements, which depend on the imaging device's pixel accuracy. In this study, we analyze the degree to which noise in the input measurements influences the performance of uLED.

\begin{figure}[!ht]
\centering
\includegraphics[width=1.0\textwidth]{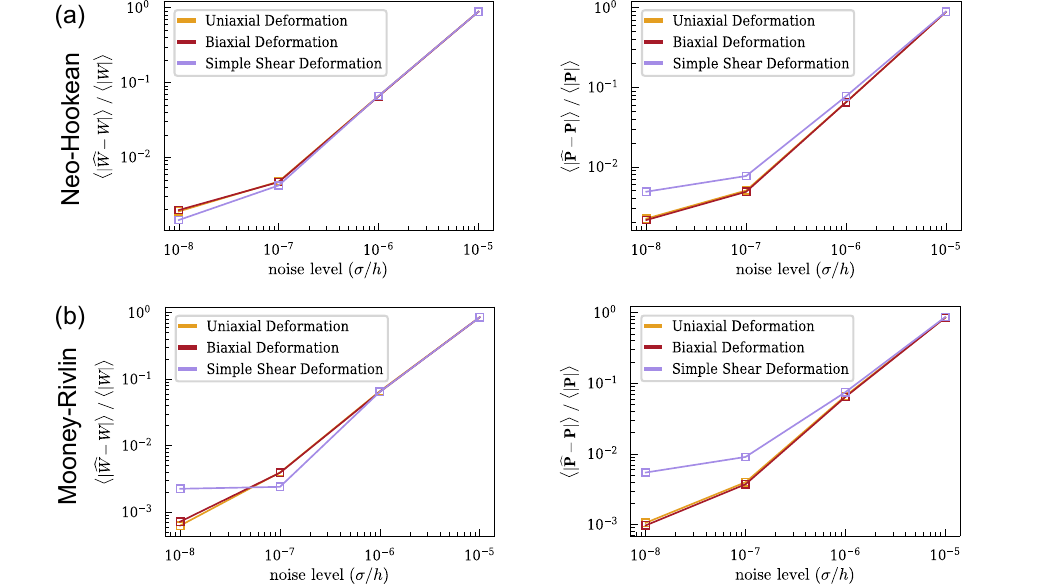}
\caption{{Convergence evaluation regarding noise level at data resolution $h_{gen}/h=1/2$.}
We compute the normalized mean absolute errors of the energy density and stress for different deformation states at a single material point, for (a) Neo-Hookean, and (b) Mooney-Rivlin materials. The predicted values are the mean of three independent experiments.}
\label{fig:exp3_noise_convergence}
\end{figure}

We generate the noisy data as follows: First, we construct the clean data with a resolution of $1/h$. Similar to \citet{thakolkaran2022nn}, we add random noise from a zero-mean normal distribution with a standard deviation $\sigma$ to the clean displacements $\vu$ to obtain the noisy displacements $\widetilde{\vu}$. In real measurements,  noise is introduced to $\vu$ due to inaccurate measurements of $\vx$  (current position) and $\mX$ (reference position). Such measurement inaccuracies scale with the data resolution, meaning that  higher data resolution results in  smaller errors. Therefore, we also add noise to our clean data that scales according to the data resolution. For our analysis, we consider four different noise levels: $\sigma =10^{-8}h, 10^{-7}h, 10^{-6}h, 10^{-5}h$. Subsequently, from these noisy displacements, we compute the noisy velocities $\dot{\widetilde{\vu}}$ and noisy accelerations $\ddot{\widetilde{\vu}}$ (see~\ref{sec:SI_data_generation} for details). This approach provides us with noisy simulation data at the data resolution $1/h$.

The introduced noise directly impacts the quality of the learned constitutive relation.  We select Neo-Hookean and Mooney-Rivlin materials as examples and examine the energy density function and stress function predicted by uLED,  trained with data affected by different noise levels $\sigma/h$ (here referred to as noisy-data-uLED). 
We compare the energy density and stress values predicted by noisy-data-uLED and the ground-truth values for multiple deformation states at \emph{a single material point}. As shown in Fig.~\ref{fig:exp3_noise_convergence}, the NMAE of the predicted energy density and stress consistently decreases with decreasing noise levels $\sigma/h$ across all considered deformations. This behavior occurs because the introduced noise prevents the constitutive relation from satisfying Eq.~\ref{eq:momentum_balance_every_node_internal}. Consequently, the learned constitutive relation deviates from the ground truth when optimizing Eq.~\ref{eq:momentum_balance_every_node_internal}. Additional analysis about the robustness to noise for other materials is reported in \ref{sec:SI_noise_levels}, showing similar performance with different levels of noise.


Therefore, we conclude that uLED's performance is quite robust to a certain level of noise. More importantly, the learned constitutive relation converges to the ground-truth constitutive relation as the noise in the measured displacements is reduced. The results suggest that uLED is potentially applicable to experimental data and that its accuracy may be further improved by increasing the precision of displacement measurements. 

\subsection{Learn constitutive relation in a dissipative setting}
\label{sec:exp4_damping}
In this experiment, we evaluate whether uLED can be applied to cases where the deformation rate also affects material behavior. To achieve this, we consider the same dynamic setting described  in Sec.~\ref{sec:validation}, but with the inclusion of energy dissipation. Consequently, the weak form of the governing equation now reads:
\begin{equation}
\label{eq:governing_equ_weak_form_with_dissipation}
     \int_{\subdomain}v_i \rho \superscriptT{t}{}{\ddot{{u}}_i} \dx + \int_{\subdomain}  v_{i,j}\superscriptT{t}{P}_{(e)ij}  \dx - \underbrace{\int_{\subdomain}  v_{i,j}\superscriptT{t}{P}_{(d)ij}  \dx}_{\text{dissipative energy}} - \int_{\subdomain}v_i \superscriptT{t}B_i \dx = 0~. 
\end{equation}
In Eq.~\ref{eq:governing_equ_weak_form_with_dissipation}, the internal strain energy $\int_{\subdomain}  v_{i,j}\superscriptT{t}{P}_{(e)ij}  \dx$ is determined by the elastic strain-stress relation $\cP_{e}$ with ${\mathbf{P}}_{(e)} = \cP_{e}(\mF)$. We use St.~Venant-Kirchhoff constitutive law for $\cP_{e}(\mF)$.  The dissipative energy $\int_{\subdomain}  v_{i,j}\superscriptT{t}{P}_{(d)ij}  \dx$ is determined by the dissipative process $\cP_{d}$ with ${\mathbf{P}}_{(d)} =\cP_{d}(\dot{\mF})$.
In this experiment, we define the dissipation function as $\cP_{d}(\dot{\mF}) = \alpha_d\cdot g(\dot{\mF})$, where $\dot{\mF} = \odv*{(\mI + \nabla_\mX  \vu)}{t}= \nabla_\mX\dot{\vu} $ and $\alpha_d$ is the dissipation coefficient. We choose $g$ to be a linear function of $\dot{\mF}$. 
 Similar to previous experiments, we set the body force to zero.


We keep the ICNN module unchanged to learn the $\cP_{e}$ but add another neural network, a simple multilayer perceptron (MLP)~\cite{bishop2006pattern}, to learn $\cP_{d}$. Importantly, we do not assume any form for $\cP_{d}$ when extending uLED to learn it. While the dissipation function is linearly related to the deformation gradient rate in the simulation, we do not provide this information to uLED. We only know that the dissipation energy depends on the deformation gradient rate $\dot{\mF}$ and  should be zero when the deformation gradient rate is zero. We enforce this constraint by using the same approach as in Sec.~\ref{sec:nn_design}. We denote the predicted $\widehat{\mP}_d$ as $\widehat{\mP}_d^{\undeformedSymbol} = \text{MLP}(\dot{\mF}^{\undeformedSymbol}; \theta_{mlp})$ by the MLP when the deformation gradient rate is zero (\textit{i.e.}, $\dot{\mF}^{\undeformedSymbol} = \mathbf{0}$). The predicted $\widehat{\mP}_d$ for any given state with $\dot{\mF}$ is $\widehat{\mP}_d(\dot{\mF}) = \text{MLP}(\dot{\mF}; \theta_{mlp}) - \widehat{\mP}_d^{\undeformedSymbol}$.
The only change in the input data compared to previous experiments (without energy dissipation) is the inclusion of the deformation gradient rate $\dot{\mF}$. 

\begin{figure}[!ht]
\centering
\includegraphics[width=1.0\textwidth]{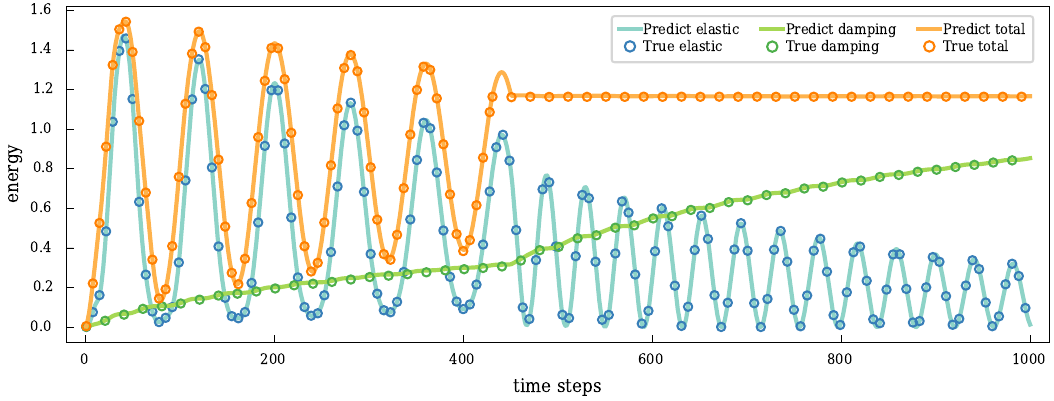}
\caption{Results for validation in a dissipative setting. 
The figure shows the evolution of energies in this dissipative system over time. The plotted predicted energies are the mean computed from three independent experiments. The standard deviation is vanishingly small.}
\label{fig:exp4_system_energies}
\end{figure}

To evaluate the learning performance, we compare the evolution of energies in this dissipative system, as shown in Fig.~\ref{fig:exp4_system_energies}. 
We observe that both the elastic energy and the accumulated dissipative energy predicted by uLED  closely align with the ground truth. This indicates that the elastic strain-stress relationship $\cP_e$ and the energy dissipation function $\cP_d$ are effectively learned. Additionally, we examine the curve of the predicted total energy, which is the sum of the elastic, dissipative, and kinetic energies, and find it to be in close agreement with the ground truth. These results suggest that uLED is potentially applicable for learning the constitutive relations in dissipative settings.

\section{Discussion}

The proposed uLED model is a powerful tool for uncovering the constitutive relation of materials. 
Its key advantage lies in the ability to infer the underlying constitutive relation solely from recorded displacements over time, without requiring prior knowledge about the constitutive relation or measurements of the boundary forces. These features make uLED particularly suitable for two challenging scenarios: 1) learning the constitutive relation of novel and unconventional materials, and 2) learning the constitutive relation in \textit{in-situ} situations where only displacements can be conveniently measured.

The advantage of uLED stems from its hybrid approach, which combines deep learning with physical laws. On the one hand, we leverage the universal approximation capability of neural networks to approximate constitutive relations without requiring any prior knowledge of the material model. On the other hand, by considering the force equilibrium for \emph{internal} nodes as the training objective, we only need to record the displacements over time for training the neural network. Stress data and boundary forces are not included in the training objective, thus eliminating the  need for  this additional information.

The learned constitutive relation is ensured to be \emph{physics-consistent} by three key design features. First, the hybrid approach inherently imposes physical constraints on the learned constitutive relation; the predicted stress naturally satisfies the conservation of linear momentum, which  is our training objective. Second, we exclusively  use strain invariants ($I_1, I_2$ for the 2D case) as  inputs to the neural network. This ensures that the learned energy density is rotation-invariant and that the learned stress is rotation-equivalent. Third, we employ the Input Convex Neural Network (ICNN) to approximate the underlying constitutive relation. With ICNN, we guarantee that the learned energy density is a convex function of the Green-Lagrangian strain, which in turn ensures local ellipticity and local material stability.

While the ICNN ensures the desired local material stability, it requires that some of its learnable parameters remain non-negative. Therefore, learning constitutive relation is formulated as a constrained optimization task (Eq.~\ref{eq:objective_function}). We utilize the projected gradient descent algorithm for training and demonstrate that ICNN can be effectively optimized to learn various constitutive relations by this optimization strategy.

When validating the proposed uLED method, we maintain a consistent  neural network architecture (\textit{e.g.}, the number of layers, the number of neurons in each layer, the activation function, etc.) across  all constitutive relations considered in the paper. Despite  this uniform  setup, we have demonstrated that uLED effectively learns different constitutive relations. However, we believe that performance can be further improved by tuning the neural network architecture for each specific constitutive relation and selecting the optimal architecture accordingly.

The main limitation of the proposed uLED method is that the constitutive relation is learned \emph{implicitly}, meaning that a black-box neural network encodes the underlying constitutive relation, and we do not obtain an explicit form of this function. This issue can potentially be addressed by applying symbolic regression (\textit{e.g.},~\cite{petersen2021deep}) to learn the explicit form of the material model. Specifically, for a given strain, the trained neural network can output the corresponding stress. Then, one can employ any symbolic regression method to discover the explicit equation relating the input strain to the corresponding stress. Notably, the field of symbolic regression is rapidly advancing, and the quality of the discovered symbolic equation will significantly depend on the chosen symbolic regression algorithm.

Although this work presents the application of uLED in 2-dimensional space, the approach can easily be extended to 3-dimensional space. The only difference for 3-dimensional space is that the strain invariants will have three values instead of two. We can use these three strain invariants as inputs to an ICNN in uLED to learn the constitutive relation in 3-dimensional space. In the future, we aim to extend the framework to encompass a broader spectrum of constitutive relations of solids, such as viscoelasticity and plasticity, by incorporating strain history and hardening phenomena through tailored neural network architectures~\cite{chen2021recurrent, li2022plasticitynet}. 
While we employ FEM to discretize the domain and approximate the displacement field, we directly minimize the loss based on the governing equation (Eq.~\ref{eq:optimization_problem}), without computing the solution using the FEM solver. This reduces computational cost and avoids difficulties in backpropagating gradients through the FEM solver. Moreover,
uLED could also be developed based on other numerical methods. Since some materials, such as granular materials, are better described by numerical frameworks other than FEM (\textit{e.g.}, material point method~\cite{yue2018hybrid, daviet2016semi}), we could consider the most suitable numerical method to solve the optimization problem in different scenarios. This flexibility in integrating different numerical methods enables uLED to be adapted to learn a wide range of materials.


\section{Conclusions}
\label{sec:conclusion}
In this research, we propose an unsupervised machine learning approach called uLED to learn the constitutive laws of hyperelastic materials. The key feature of our approach is that it learns \emph{physics-consistent} constitutive relations without requiring any information about the stress labels or boundary forces as input. These advantages are achieved by leveraging dynamic, rather than static, observations of the displacement field. 
We have demonstrated that the proposed method can be applied to learn a diverse range of materials solely using temporal displacement data.
We have also demonstrated that partial observations of the displacement field are sufficient to learn the constitutive relation.
Moreover, once uLED is trained based on the observed dynamics of one sample, the trained model can be directly applied to other samples of the same material with different geometries. This is because the constitutive relation learned by uLED is independent of the specific material sample used for training. uLED demonstrates excellent generalizability, effectively learning constitutive relations for various materials. 
Furthermore, we have demonstrated that the learned constitutive relation is robust to significant levels of noise and converges to the ground truth as data resolution increases. Finally, the proposed approach, which builds on observations of dynamic displacement fields, is particularly powerful for \textit{in-situ} applications. This includes inferring the material model of biological tissues and strain-rate-dependent materials, providing an opportunity to discover accurate constitutive relations in complex real-life scenarios. In the future, we aim to extend uLED to real applications in 3-dimensional space.

\section{Acknowledgements}
This project has been funded by ETH Grant no. ETH-12 21-1.

\section{Code availability}
The FEM simulation is implemented with FEniCS~\cite{alnaes2015fenics} (version 2019).
The implementation of the proposed uLED method is based on PyTorch~\cite{paszke2019pytorch}. 
The source code is available on Gitlab: \url{https://gitlab.ethz.ch/cmbm-public/papers-supp-info/2024/learning-physics-consistent-material-behavior-without-prior-knowledge}.

\bibliographystyle{elsarticle-num-names} 
\bibliography{ref}





\appendix

\section{ICNN model configuration}
\label{sec:SI_ICNN_configuration}

To ensure that the learned energy density is a convex function of strain invariants, we use an ICNN~\cite{amos2017input} as the neural network model. 
The layer-wise propagation rule of ICNN is:
\begin{equation*}
    \vz_{i+1} = g_i(\vz_i W_i^{(z)} + \vx W_i^{(x)} + b_i), \quad \text{for } i = 0, 1, \ldots
\end{equation*}
where $\vz_{i+1}$ is the intermediate latent variable at layer $i+1$ ($\vz_0 \equiv 0$). $\vx$ is the input (strain invariants in this paper).  $W_i^{(z)}$ are the non-negative learnable weights at the $i$-th layer ($W_0^{(z)}\equiv 0$). $W_i^{(x)}$ and $b_i$ are the standard  learnable weights and biases without the non-negative restriction at the $i$-th layer. For convenience, we refer to the non-negative weights $W_i^{(z)}$ as $\theta_{i}^{\text{(cvx)}}$ and the standard weights $W_i^{(x)}$ and bias $b_i$ at the $i$-th layer as $\theta_{i}^{\text{(fc)}}$. The terms $\theta^{\text{(cvx)}}$ and $\theta^{\text{(fc)}}$ refer to all non-negative and standard  learnable parameters, respectively. 

In this work, we use a 5-layer ICNN to learn the constitutive relation of all considered materials. The neural network architecture is shown in the following table:

\begin{table}[h!] 
  \centering
  \caption{Layer weights and bias configuration.}
  \label{table:SI_ICNN_architecture}
  \resizebox{1\textwidth}{!}{%
  \begin{tblr}{colspec={Q[m,c,3cm] Q[m,c,12cm]},row{1}={4ex}, row{2-6}={5ex}}
    \toprule[1.5pt]
    layer & parameter configuration  \\
    \midrule
    layer 0 & $W_0^{(z)} \equiv 0$, $\quad W_0^{(x)}$ shape: [2, 64], $\quad b_0$ shape: [64]  \\
    \hline
    layer 1 & $W_1^{(z)}$ shape: [64, 64], $W_1^{(x)}$ shape: [2, 64], $b_1$ shape: [64]  \\
    \hline
    layer 2 & $W_2^{(z)}$ shape: [64, 64], $W_2^{(x)}$ shape: [2, 64], $b_2$ shape: [64]  \\
    \hline
    layer 3 & $W_3^{(z)}$ shape: [64, 64], $W_3^{(x)}$ shape: [2, 64], $b_3$ shape: [64]  \\
    \hline
    layer 4 & $W_4^{(z)}$ shape: [64, 1], $W_4^{(x)}$ shape: [2, 1], $b_4$ shape: [1]  \\
    \bottomrule[1.5pt]
    \end{tblr}
    }
\end{table}

Additionally, ICNN requires the non-linear activation functions $g_i$ to be convex and non-decreasing. In this work, we use the Exponential Linear Unit (ELU)~\cite{clevert2016fast} as the activation function for all layers, except for the output layer.

\section{Numerical computation of the optimization problem (Eq.~\ref{eq:optimization_problem})}
\label{sec:background}

To solve the optimization problem defined in Eq.~\ref{eq:optimization_problem}, we use the FEM to approximate the $\superscriptT{t}\functional_i(\theta; v_i)$, as defined in Eq.~\ref{eq:governing_equ_weak_form}. In this work, we use triangular elements to discretize the reference domain and use linear basis functions $\{N^a(\mX)\}$, where $a \in \mathcal{C}_n$ nodes (with $\mathcal{C}_n$ being  the set of internal nodes), to approximate the trial and the test function, as indicated by Eq.~\ref{eq:approximate_u_v}:

\begin{equation}
\label{eq:approximate_u_v}
\superscriptT{t}\vu(\mX) \approx \sum_{a=1}^{n_n}N^a(\mX)\superscriptT{t}\vu^{a}\, , \quad
\vv(\mX) \approx \sum_{a=1}^{n_n}N^a(\mX)\vv^a
\end{equation}
where $\superscriptT{t}\vu^{a}$ and $\vv^a$ are the values of the trial function and test function of node $a$ at time t.

By approximating the $\vu$ and $\vv$ in Eq.~\ref{eq:governing_equ_weak_form} using Eq.~\ref{eq:approximate_u_v}, the balance equation for these internal nodes reduces to (Eq.~\ref{eq:momentum_balance_every_node_internal})
\begin{equation}
\label{eq:appendix_momentum_balance_internal_node}
\begin{split}
    &\forall t, \ \forall i:\quad \quad \sum_{a=1}^{n_n} v_i^a \superscriptT{t}f_i^{a}(\theta) = 0 \, , \\
    &\superscriptT{t}f_i^{a}(\theta) = \sum_{\text{el} \in \text{supp}(N^a)} \left[\int_{\text{el}}N^a \rho N^b \dx \cdot \superscriptT{t}{\ddot{u}}_i^{b} + \int_{\text{el}}  N^a_{,j} \superscriptT{t}\widehat{P}_{ij} \dx - \int_{\text{el}}N^a \superscriptT{t}B_i \dx \right]
\end{split}
\end{equation}
where $\text{supp}(N^a)$ is the set of elements for which $N^a$ is non-zero. 
Since the test function can have arbitrary values for internal nodes, the function $f_i^a(\theta)$ in Eq.~\ref{eq:appendix_momentum_balance_internal_node} must be zero for these internal nodes, \textit{i.e.}, 
\begin{equation}
     \forall i\in d, \forall t:\quad \superscriptT{t}f_i^{a}(\theta) = 0
\end{equation}
Therefore, we derive the optimization objective as shown in Eq.~\ref{eq:objective_function}. In the following, we explain how to numerically compute each term in $\superscriptT{t}f_i^{a}(\theta)$.
Since  we use linear basis functions in this paper, the term $\int_{\text{el}}N^a \rho N^b \dx$ can be computed analytically. Additionally, $N^a_{,j}$ and $\superscriptT{t}{\mF}$ are constant within each element because the basis functions are linear. Consequently, the $\superscriptT{t}\widehat{\mP} = \cPapprox(\superscriptT{t}\mF; \theta)$ is constant within each element. Therefore, the term $\int_{\text{el}} N^a_{,j} \superscriptT{t}\widehat{P}_{ij} \dx$ can be computed by multiplying  $ N^a_{,j}$, $\superscriptT{t}\widehat{P}_{ij}$, and the volume of the element. We do not consider body forces, so the term $\int_{\text{el}}N^a \superscriptT{t}B_i \dx$ is zero for each element. 


\section{Details in uLED training}
\label{sec:SI_training_details}
We randomly initialize the parameters in uLED using the Xavier initialization algorithm~\cite{glorot2010understanding}. We use the Adam optimizer~\cite{kingma2014adam} with a learning rate of 5E-4 to optimize all learnable parameters. We split the input along the time dimension to mini-batches, with the batch size chosen to avoid exceeding the GPU memory capacity. The maximum number of training epochs is set to 300 for all materials to ensure the proposed ML model achieves good accuracy (see Fig.~\ref{fig:exp1_train_curves}). To ensure  $\theta^{\text{(cvx)}}$ remains non-negative, we project the negative values in $\theta^{\text{(cvx)}}$ to zero after each gradient descent step.
The pseudo-code for the training procedure is summarized in Algorithm~\ref{algorithm:uLED_training}.
The experiments are conducted on a server equipped with four RTX 3090 GPUs.

\begin{algorithm}[!h]
\setstretch{1.2}
\SetAlgoLined
\SetAlgoLined\DontPrintSemicolon
\SetKwFunction{algo}{algo}\SetKwFunction{projection}{projection}\SetKwFunction{absolute}{absolute}
\SetKwComment{Comment}{\#\ }{}
\caption{uLED training procedure}
\label{algorithm:uLED_training}
Prepare time-invariant inputs: $\rho, \{N^a\}_{a=1,\ldots,n_n}, \{N_{,j}^a\}_{a=1,\ldots,n_n}$\;
Prepare displacements at data resolution $1/h$: $\{\superscriptT{t}\vu^{a}\}_{a=1,\ldots,n_n}^{t=1,\ldots,T}$, $\{\superscriptT{t}\ddot{\vu}^{a}\}_{a=1,\ldots,n_n}^{t=1,\ldots,T}$ \;
Randomly initialize $\theta^{\text{(cvx)}}$ as $\theta^{\text{(cvx)}, \text{now}}$,  $\theta^{\text{(fc)}}$ as $\theta^{\text{(fc)}, \text{now}}$ \;
$\theta^{\text{(cvx)}, \text{now}} \gets$  \absolute($\theta^{\text{(cvx)}, \text{now}}$)\;
\Repeat{converge or reach max training epochs}{
Split $\{\superscriptT{t}\vu^{a}\}_{a=1,\ldots,n_n}^{t=1,\ldots,T}$, $\{\superscriptT{t}\ddot{\vu}^{a}\}_{a=1,\ldots,n_n}^{t=1,\ldots,T}$ to mini-batches along time steps\;
\Comment*[l]{If necessary, split along spatial domain as well}

\For{each batch }{ 
    \nonl \Comment*[l]{Compute the batch loss}
    \nl Compute $\mathcal{L}(\theta^{\text{(cvx)}}, \theta^{\text{(fc)}})$ by Eq.~\ref{eq:objective_function} \;
    
    \nonl \Comment*[l]{Optimize parameters, using Adam as the optimizer}
        
    \nl $\theta^{\text{(fc)}, \text{new}} \gets$  Optimizer.step $\left(\nabla_{\theta^{\text{(fc)}}} \mathcal{L}(\theta^{\text{(cvx)}}, \theta^{\text{(fc)}}),  \theta^{\text{(fc)}, \text{now}} \right)$\;

    \nl $\theta^{\text{(cvx)}, \text{new}} \gets$  Optimizer.step $\left(\nabla_{\theta^{\text{(cvx)}}} \mathcal{L}(\theta^{\text{(cvx)}}, \theta^{\text{(fc)}}), \theta^{\text{(cvx)}, \text{now}} \right)$\;

    \nl $\theta^{\text{(cvx)}, \text{new}} \gets$  \projection($\theta^{\text{(cvx)}, \text{new}}$)\;

    \nonl \Comment*[l]{Update the current estimation of $\theta^{\text{(cvx)}}$ and $\theta^{\text{(fc)}}$}
    
    \nl $\theta^{\text{(cvx)}, \text{now}} \gets \theta^{\text{(cvx)}, \text{new}}$\;
    \nl $\theta^{\text{(fc)}, \text{now}} \gets \theta^{\text{(fc)}, \text{new}}$\; 
}
}
\nonl \;
\setcounter{AlgoLine}{0}
\SetKwProg{myproc}{Function}{}{}
\myproc{\absolute{$\theta$}}{
\For{each element $\theta_{i,j}$ in $\theta$ }{
\uIf{$\theta_{i,j}$ < 0}{
$\theta_{i,j}$ = $-\theta_{i,j}$ \;
}
}
\KwRet $\theta$\;}

\nonl \;
\setcounter{AlgoLine}{0}
\SetKwProg{myproc}{Function}{}{}
\myproc{\projection{$\theta$}}{
\For{each element $\theta_{i,j}$ in $\theta$ }{
\uIf{$\theta_{i,j}$ < 0}{
$\theta_{i,j}$ = 0 \;
}
}
\KwRet $\theta$\;}

\end{algorithm}


\section{FEM simulation settings}
\label{sec:SI_data_generation}
We use dimensionless units for all simulations, as the
machine-learning algorithm is not designed for any specific scale. For  the simulations, we set Young's module to 10,000, the Possion's ratio to 0.3, and the density $\rho$ to 1.0. Correspondingly, this results in $\lambda = 5769.2$ (Lam\'e's first parameter), $\mu = 3846.2$ (Lam\'e's second parameter), and $K = 8333.3$ (bulk modulus).

All simulations run 1,500 time steps with a time step size of 0.002. 
We use the implicit Newmark method to update the dynamics for the FEM simulation. We use triangle mesh for the FEM simulation, with element size of $h_{gen} = 0.001$. The velocity and acceleration are updated as follows:
\begin{equation}
\label{eq:newmark_update}
\begin{split}
    &\superscriptT{t+1}\ddot{\vu} = \frac{\superscriptT{t+1}\vu - \superscriptT{t}\vu - \Delta t \cdot \superscriptT{t}\dot{\vu}}{\beta \Delta t^2} - \frac{1 - 2\beta}{2 \beta}\superscriptT{t}\ddot{\vu}\\
    & \superscriptT{t+1}\dot{\vu} =  \superscriptT{t}\dot{\vu} + \Delta t ((1-\gamma) \superscriptT{t}\ddot{\vu} + \gamma\cdot \superscriptT{t+1}\ddot{\vu})
\end{split}
\end{equation}
We denote  the displacements at the next time step $t+1$ as $\superscriptT{t+1}\vu$ and the displacements at the current time step $t$ as $\superscriptT{t}\vu$  (similarly  for velocities and accelerations). We set $\beta = \frac{1}{4}$ and $\gamma = \frac{1}{2}$ to ensure  the simulation unconditionally stable.

In Sec.~\ref{sec:exp3_noise} where we generate noisy data, we impose noise
on the sampled nodal displacements at the resolution $1/h$ using  the following equation:
\begin{equation*}
    \superscriptT{t}{\widetilde{u}}_{i}^{a} \leftarrow \superscriptT{t}{u}_{i}^{a} + \delta u
\end{equation*}
where $\superscriptT{t}{\widetilde{u}}_{i}^{a}$ is the $i$-th dimension of the noisy displacement of node $a$ at time $t$, and $\delta u \sim \mathcal{N}(0, \sigma^2)$ is a random number sampled independently from a zero-mean normal distribution with standard deviation $\sigma$. After obtaining  the noisy sampled nodal displacements $\{\superscriptT{t}\widetilde{\vu}\}_{t=1,\ldots,T}$, we generate the noisy velocities and accelerations by using the noisy displacements in the Newmark update scheme (Eq.~\ref{eq:newmark_update}).

\section{Details to construct the uLED input from FEM simulation}
\label{sec:construct_input_from_FEM}
\begin{figure}[!ht]
\centering
\includegraphics[width=1.0\textwidth]{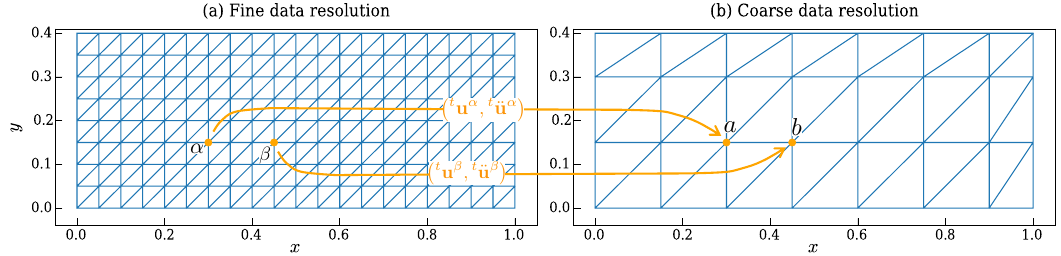}
\caption{Illustration of the mesh coarse-graining and nodal dynamics interpolation. (a) The fine mesh with a mesh size of 0.05. (b) A coarse mesh created by taking every third node ($k=3$) along each dimension. We only simulate the nodal dynamics (displacements, velocities, accelerations) on the fine mesh. The nodal dynamics on the coarse-grained mesh are directly interpolated from the corresponding nodes on the fine data resolution. For example, the node marked by `a' with position (0.3, 0.15) on the coarse mesh has the same dynamics as the node marked by `$\alpha$' at the same position on the fine mesh. Note that this figure is for illustration purposes. We use a smaller mesh size ($h_{gen}=0.001$) for the fine data resolution in this paper.}
\label{fig:mesh_discretization}
\end{figure}

The FEM simulation generates nodal dynamics, including displacements and accelerations at data resolution of $1/h_{gen}$. To construct the input for training uLED at the coarser data resolution $1/h$ ($1/h < 1/h_{gen}$), we create a coarse mesh by selecting  every $k$-th node of the fine mesh (at resolution $1/h_{gen}$) along each spatial dimension, as illustrated  in Fig.~\ref{fig:mesh_discretization}. We vary  $k$ from $1$ to $10$ to create different levels of coarse-grained meshes. When $k$ is $1$, the new mesh is identical to the fine mesh, resulting in no discretization error. Conversely, when $k$ is $10$,  a significant  discretization error is introduced. The nodal displacements and accelerations in each coarse-grained mesh are directly interpolated from the recorded nodal displacements and accelerations in the fine mesh (see Fig.~\ref{fig:mesh_discretization}). 
Throughout the main manuscript (except in Sec.~\ref{sec:exp1_discretization}, where we study the effect of different data resolutions), we set $h_{gen}/h  = 1/2$, \textit{i.e.}, $k=2$.

When constructing the input for uLED, we sample every 7-th time step from the 1500 time steps for each simulation, resulting in  214 time steps for training and validation. Among these 214 time steps, we use the first 80\% for training and the remaining 20\%  for validation. The best-trained model is selected based on the objective function (mean absolute error of nodal force balance, Eq.~\ref{eq:objective_function}) evaluated on the validation dataset. Notably, we do not use any information about the ground-truth stress labels.

\section{Visualization of the learned energy density  and stress functions of different materials}
\label{sec:SI_W_P_along_Fpaths}
\begin{figure}[!h]
\centering
\includegraphics[width=1.0\textwidth]{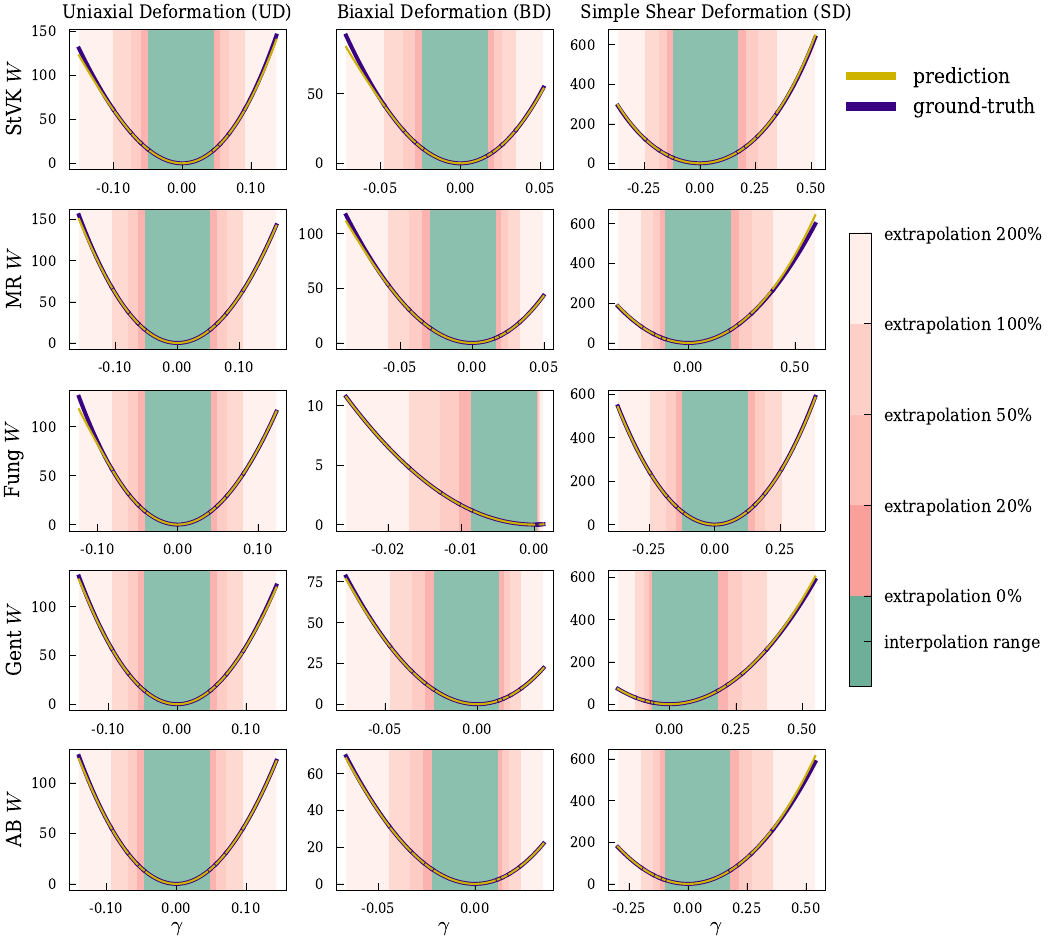}
\caption{
Validation of the predicted energy density of various materials. Curves show the predicted and ground-truth energy density for different deformation states \emph{at a single material point}, in which the prediction is the mean of three independent experiments.}
\label{fig:exp1_preicted_W}
\end{figure}

\begin{figure}[!h]
\centering
\includegraphics[width=1.0\textwidth]{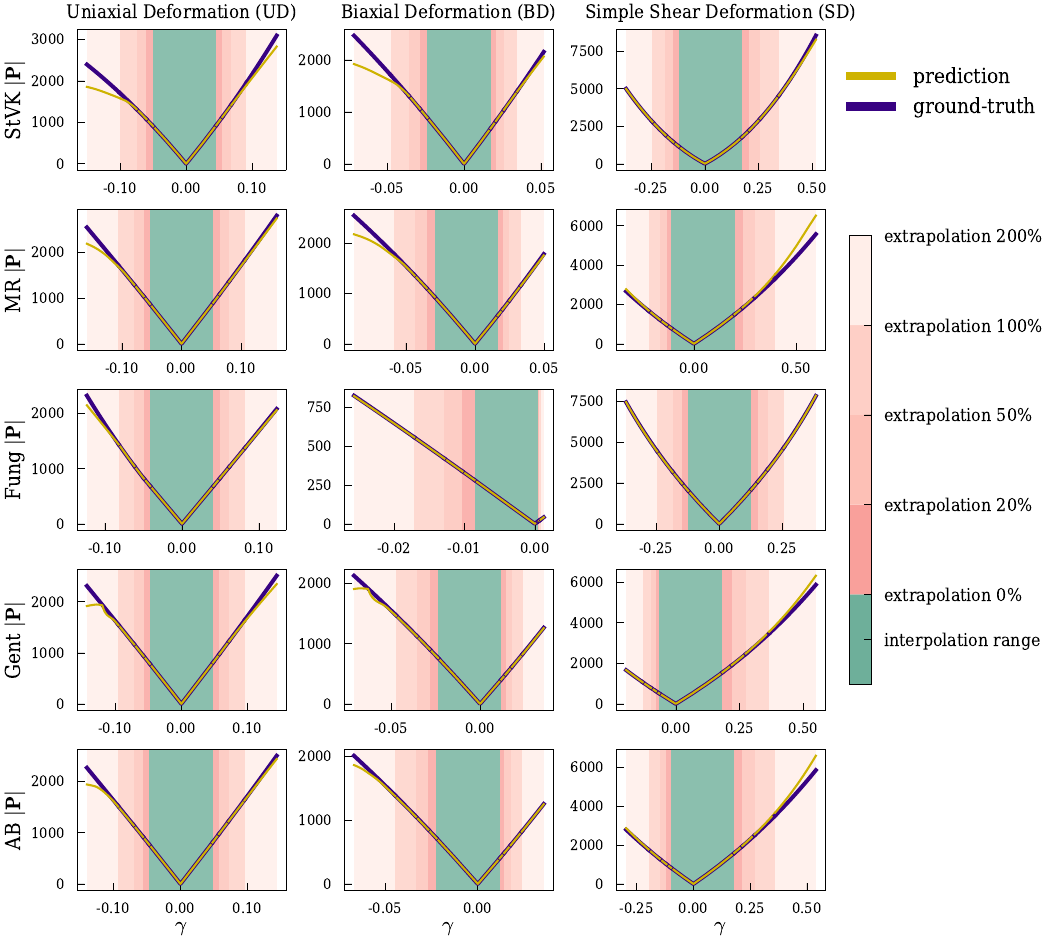}
\caption{
Validation of the predicted stress of various materials.
Curves show the predicted and ground-truth stress for different deformation states \emph{at a single material point}, in which the prediction is the mean of three independent experiments.}
\label{fig:exp1_preicted_P}
\end{figure}

In Sec.~\ref{sec:NH_showcase}, we analyzed the curves of the predicted and ground-truth energy densities for the Neo-Hookean material across different deformation states (see Eq.~\ref{eq:F_path}) at a single material point. Here, we extend this comparison to other materials (see Table~\ref{table:hyperelatic_models}), by comparing the curves of the predicted energy density and stress to the ground truth, as shown in Fig.~\ref{fig:exp1_preicted_W} and Fig.~\ref{fig:exp1_preicted_P}. The results align with the analysis for the Neo-Hookean material in Sec.~\ref{sec:exp2_various_materials}.


\section{Training curves}
\label{sec:SI_training_curves}
\begin{figure}[!ht]
\centering
\includegraphics[width=1.0\textwidth]{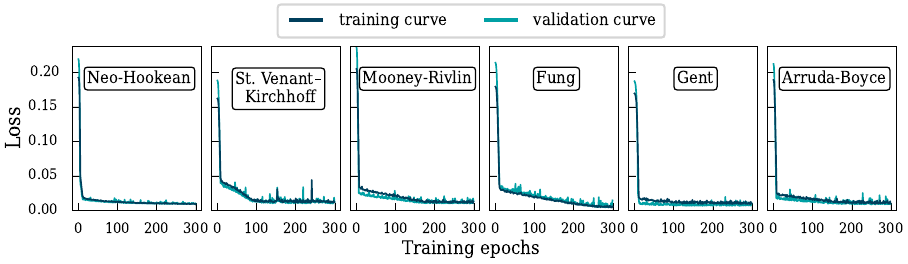}
\caption{{Training curves of uLED for different materials.} The training loss and validation loss are computed by Eq.~\ref{eq:objective_function} using the training data and validation data. All materials are trained with 300 epochs. Each experiment of training uLED is repeated three times with randomly initialized parameters in ICNN -- one arbitrary example is shown.}
\label{fig:exp1_train_curves}
\end{figure}

To examine the efficiency of the optimization strategy for various material constitutive laws, we analyze the training loss and validation loss during the training process. For all considered materials, both the training loss and validation loss decrease significantly during the training process (see Fig.~\ref{fig:exp1_train_curves}), demonstrating that the proposed optimization strategy efficiently trains uLED to learn the material models. Additionally, we note that the validation and training loss are similar throughout the training process for all considered material laws, indicating that there is generally no overfitting in training uLED.

\section{Visualization of learned energy density and stress fields at different data resolutions}
\label{sec:SI_convergence_data_resolution}

\begin{figure}[!ht]
\centering
\includegraphics[width=1.0\textwidth]{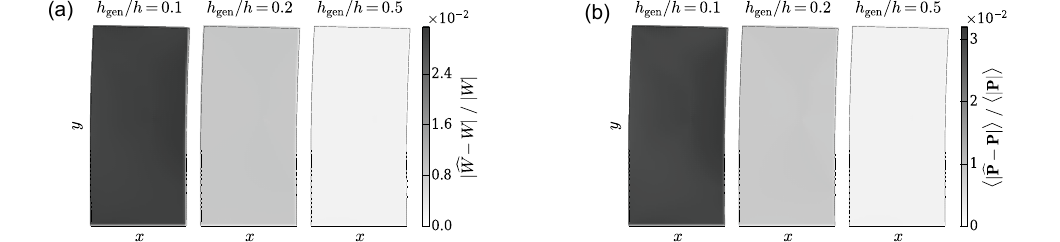}
\caption{Convergence evaluation regarding data resolution for Neo-Hookean material. We train uLED with different data resolutions and compare the predicted  (a) energy density fields and (b) stress fields to the ground-truth. The predicted values are the mean of three independent experiments.}
\label{fig:W_P_fields_data_resolution}
\end{figure}

We analyze the predicted energy density and stress fields by uLED trained with different data resolutions. To this end, we randomly select a deformation state used in training (the one shown in Fig.~\ref{fig:exp0_W_P_field_same}) and apply uLED,  trained with different data resolutions, to predict the energy density for this deformation state. Results are reported in Fig.~\ref{fig:W_P_fields_data_resolution}, showing the predicted energy density and stress fields get closer to the ground-truth with higher data resolutions. 

\section{Visualization of learned energy density and stress functions with different noise levels}
\label{sec:SI_noise_levels}
\begin{figure}[!ht]
\centering
\includegraphics[width=1.0\textwidth]{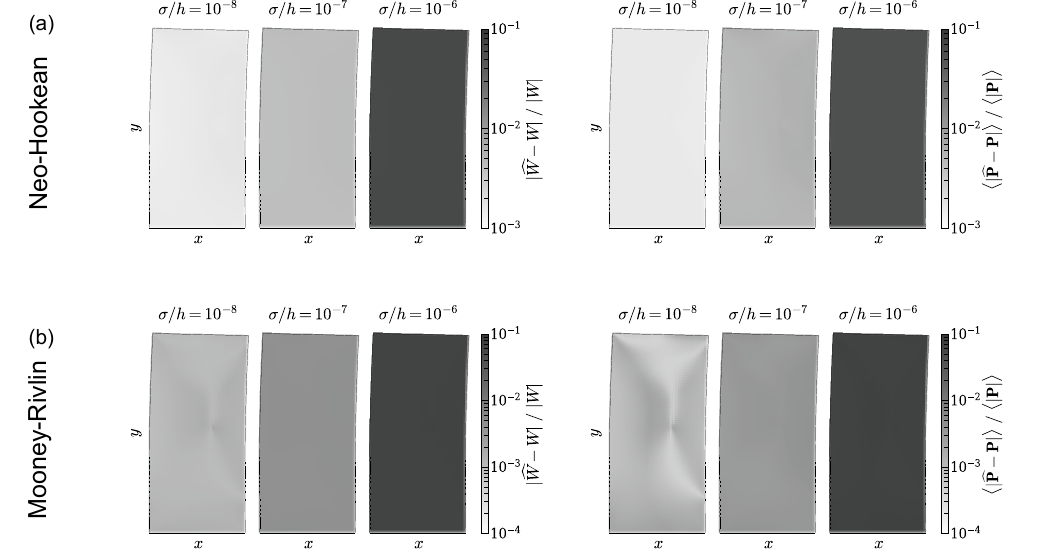}
\caption{The error in predicted energy density fields and stress fields at different noise levels. We randomly select a deformation state used in training (the reference case in Fig.~\ref{fig:exp0_W_P_field_same}~a) and compare the predicted energy density fields and stress fields by the noisy-data-uLED to the ground-truth. The predicted values are the mean from three independent experiments.
}
\label{fig:W_P_fields_noise_dta}
\end{figure}

\begin{figure}[!ht]
\centering
\includegraphics[width=1.0\textwidth]{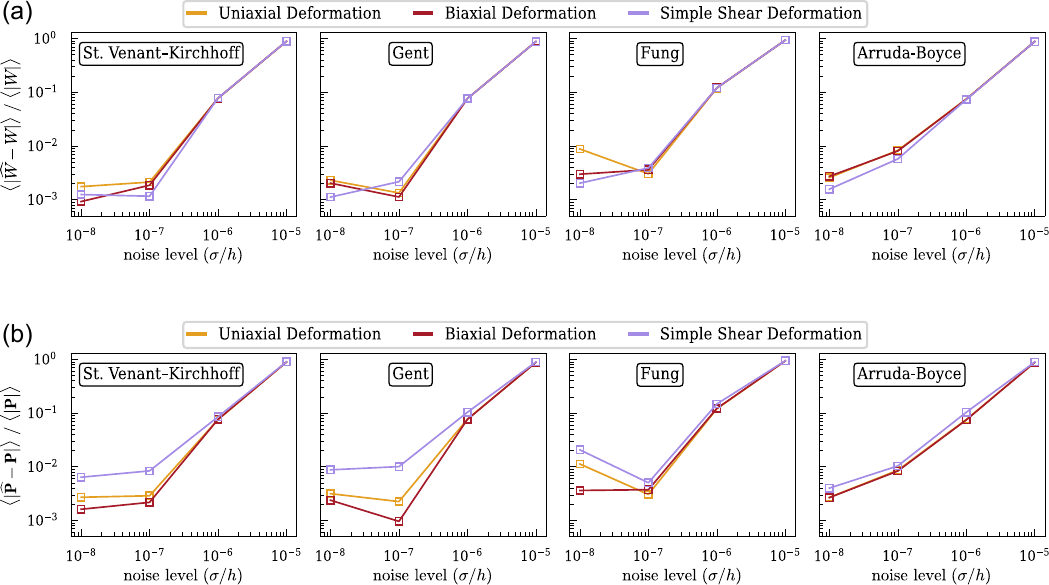}
\caption{Convergence evaluation regarding noise levels for various materials. The curves show the NMAE of the energy density and stress for different deformation states at a single material point. The prediction is the mean of three independent experiments.}
\label{fig:noisy_error_convergency_combined}
\end{figure}

In Sec.~\ref{sec:exp3_noise}, we report the NMAE of the predicted energy density and stress of uLED trained with different noise levels for Neo-Hookean and Mooney-Rivlin materials. In Fig.~\ref{fig:W_P_fields_noise_dta}, we compare the corresponding energy density fields and stress fields of the noisy-data-uLED to the ground-truth. It demonstrates that the difference between the prediction and the ground-truth decreases with lower noise levels ($\sigma/h$). Further, we extend the convergence evaluation regarding noisy levels to other materials listed in Table~\ref{table:hyperelatic_models} (see Fig.~\ref{fig:noisy_error_convergency_combined}). 

\section{Comparison of training strategies}
\label{sec:SI_training_strategy}
\begin{figure}[!ht]
\centering
\includegraphics[width=1.0\textwidth]{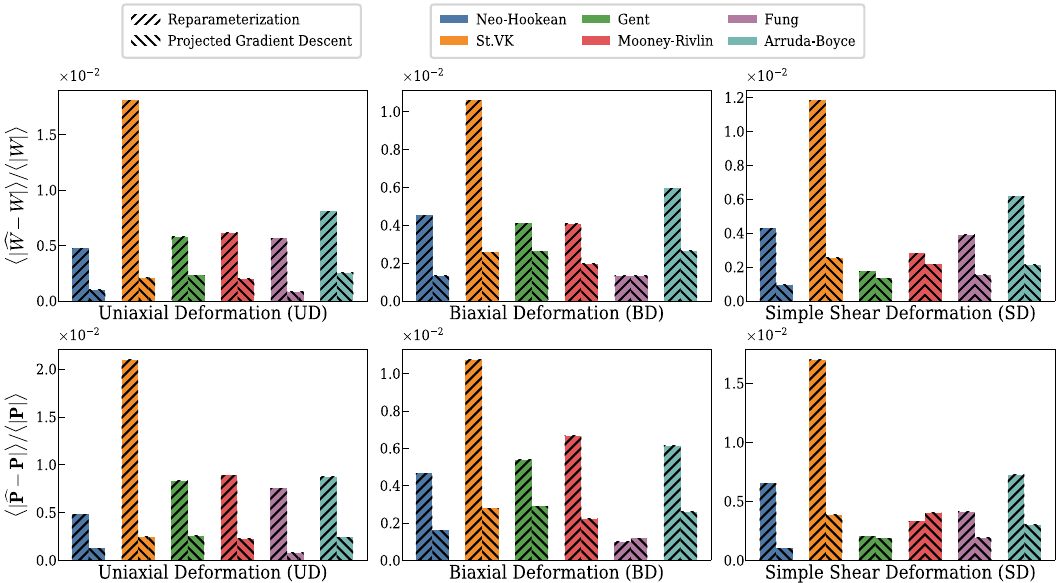}
\caption{{Comparison of two possible training strategies.} Reparameterization is used by NN-EUCLID. Projected gradient descent is used by uLED. The predicted values in the reported NMAE are the mean of three independent experiments.}
\label{fig:compare_EUCLID_uLED_W}
\end{figure}

We compare the performance of two training strategies: reparameterization used in NN-EUCLID~\cite{thakolkaran2022nn} and the projected gradient descent used in uLED. To this end, we keep the NN-EUCLID model unchanged but use strain invariants as the input. We train NN-EUCLID on the same data used for training uLED. Hence, the experimental setting is the same for NN-EUCLID and uLED, allowing us to compare two different strategies that ensure the non-negativity constraint of ICNN. As shown in Fig.~\ref{fig:compare_EUCLID_uLED_W}, the projected gradient descent approach consistently performs better than the parameterization approach on the considered materials.

\end{document}